\documentclass[12pt,draftclsnofoot,letterpaper,onecolumn,romanappendices]{IEEEtran}
\usepackage[dvips]{graphicx}
\usepackage{cite}
\usepackage{epstopdf}
\usepackage{epsfig}
\usepackage{amsthm}
\usepackage{amsmath,amssymb,amsfonts,amstext,amsbsy,amsopn,dsfont}
\usepackage{cases}
\usepackage{sublabel}
\usepackage{array}

\newtheorem{theorem}{\textbf{Theorem}}

\newtheorem{lemma}{\textbf{Lemma}}
\newtheorem{remark}{\textbf{Remark}}
\newtheorem{definition}{\textbf{Definition}}

\newcommand{\defn}{\triangleq}
\newcommand{\dif}{\textmd{d}}

\begin{document}

\title{Optimal Discrete Power Control in Poisson-Clustered Ad Hoc Networks}

\author{Chun-Hung Liu, Beiyu Rong, and Shuguang Cui\\
\thanks{C.-H. Liu is with the Department of Electrical and Computer Engineering at National Chiao Tung University, Hsinchu, Taiwan (Email: chungliu@nctu.edu.tw). B. Rong is with Marvell Semiconductor Inc., Santa Clara, CA USA (Email: beiyurong@gmail.com).  S. Cui is with the Department of Electical and Computer Engineering at Texas A\&M University, College Station, TX USA (Email: cui@ece.tamu.edu). Dr. Liu is the contact author. Manuscript Date: \today.}
}

\maketitle

\begin{abstract}
Power control in a digital handset is practically implemented in a discrete fashion and usually such a discrete power control (DPC) scheme is suboptimal. In this paper, we first show that in a Poison-distributed ad hoc network, if DPC is properly designed with a certain condition satisfied, it can strictly work better than constant power control (i.e. no power control)  in terms of average signal-to-interference ratio, outage probability and spatial reuse. This motivates us to propose an $N$-layer DPC scheme in a wireless clustered ad hoc network, where transmitters and their intended receivers in circular clusters are characterized by a Poisson cluster process (PCP) on the plane $\mathbb{R}^2$. The cluster of each transmitter is tessellated into $N$-layer annuli with transmit power $P_i$ adopted if the intended receiver is located at the $i$-th layer. Two performance metrics of transmission capacity (TC) and outage-free spatial reuse factor are redefined based on the $N$-layer DPC. The outage probability of each layer in a cluster is characterized and used to derive the optimal power scaling law $P_i\in\Theta\left(\eta_i^{-\frac{\alpha}{2}}\right)$, with $\eta_i$ the probability of selecting power $P_i$ and $\alpha$ the path loss exponent. Moreover, the specific design approaches to optimize $P_i$ and $N$ based on $\eta_i$ are also discussed. Simulation results indicate that the proposed optimal $N$-layer DPC significantly outperforms other existing power control schemes in terms of TC and spatial reuse.
\end{abstract}

\section{Introduction}
Power control is especially crucial in a large-scale multiuser wireless network where interference is the main limiting factor in achieving high network throughput. A large volume of work, led by the pioneer results in \cite{JZANDER93,SAGRVDJGJZ93,JZANDER92,GFZM93}, has contributed to the design of optimal centralized or distributed power control schemes that could provide certain quality of service (QoS). A general framework for power control was thoroughly examined in \cite{RDYATES95} for a broad class of systems, where it is shown that if the interference function is standard, a distributed and iterative (continuous) power control algorithm converges to the minimum power solution. Although such continuous power control schemes are technically sound, they have to be discretized in practice since transmit power in a digital handset can only be updated at discrete levels \cite{MAZRJZ98}. For instance, the downlink and uplink transmit power in an IS-95 system may vary from 12 to 85 dB at steps of 0.5 dB \cite{Qualcomm92}. As such, how to design and implement discrete power control in wireless communication systems is always a key problem.

In an ad hoc network, a discrete power control (DPC) scheme is preferable to be developed in a distributed fashion to reduce control overhead, which usually results in suboptimal schemes, especially when the network size is large. In recent years, applying Poisson point process (PPP) to modeling random node locations in large-scale networks has been shown to be a valid and analytically tractable approach \cite{WeberTC,FBBBPM06,MHJGAFBODMF10}. However, the power control problem in such a framework may not be completely tractable, since the complex distribution of interference exacerbates the analyses of outage probability, network throughput, etc.. In this paper, we aim at developing a simple and tractable DPC scheme in such a PPP-based ad hoc networking frame. More generally, we consider a Poisson cluster process (PCP)\footnote{The phenomena of PCP-based node distribution can be observed in many different kinds of wireless networks, such as clustered sensor networks, mobile ad hoc networks, small cell and heterogeneous cellular networks in a large city, etc. \cite{OYMKSR06,JYYPHJC06,CHLJGA11,RKGMH09}.} to model the distributions of transmitters and receivers in a clustered ad hoc network: Transmitters form a homogeneous PPP of intensity $\lambda$, and each of them is associated with a random number of receivers in a circular cluster that is tessellated into $N$-layer annuli. 

\subsection{Previous Work}
Representative literatures on distributed power control in wireless ad hoc networks can be found in \cite{SARHKSVKSKD01,TEAE04,CWSKKL05,VKPRK05}, which usually are not designed for discrete implementation. A distributed DPC scheme cannot be simply realized by discreting a continuous distributed power control scheme, since such obtained DPC schemes may not retain the convergence and uniqueness properties \cite{MAZRJZ98}. Therefore, DPC needs its own problem formulation and analysis. For example, in \cite{SLKZRJZ99} the authors studied the joint optimization problem of discrete power and rate control. The problem of minimizing the sum power subject to signal-to-noise ratio constraints was considered in \cite{CWUDPB01}. Meanwhile, game-theoretic distributed DPC formulation is popular. In \cite{MHRPMPEC04}, a game-theoretic formulation for non-cooperative power control with discrete power levels and channel fading states is proposed, while \cite{YXRC08} formulated the distributed DPC problem as a utility-based $N$-person nonzero-sum game with a stochastic iterative process. Reference \cite{EAKAIMGMBJPAS09} investigates the dynamic discrete power control scheme in uplink cellular networks in which the transmit power level of a user is chosen based on the available channel state information. Although the above schemes succeed in achieving a certain level of power optimality, they are unable to provide tractable analytical performance metrics, such as outage probability, network throughput, etc.. In addition, their results are mainly restricted to small network topologies, such that useful insights on the behaviors of large-scale networks are hardly perceived.

In the framework of Poisson-distributed ad hoc networks, a few heuristic power control algorithms have been studied, with the popular approach of combating the fading effect. For example, channel inversion power control studied in \cite{SWJGANJ07} sets the transmit power as the inverse of the channel gain between a transmitter and its intended receiver. For some fading distributions like Rayleigh fading, the inverse channel gain can be infinitely large, which is infeasible to implement. Another similar power control scheme, called fractional power control, is a modified version of channel-inversion power control and its idea is to make the transmit power to be a partially inverse function of the fading channel gain \cite{NJSPWJGA08}. These channel-aware power schemes require the knowledge of instantaneous fading gains at every time slot and thus their performance may significantly degrade when erroneous channel estimation happens. The ALOHA-type random on-off power control policies and delay-optimal power control policies in a Poisson-distributed wireless network are studied in \cite{XZMH0912} and \cite{XZMH1012}, respectively. All these prior literatures on power control in Poisson-distributed wireless networks are not discrete and thus implementing them in a discrete way certainly undermines their original idea of combating/canceling fading. In addition, the signal reception quality could be remarkably affected by the transmission distance, which means,  an efficient DPC scheme should be of the distance-aware nature. This is the core idea of the proposed $N$-layer DPC scheme in this paper.

\subsection{Contributions}
Our first contribution is to identify under what conditions the DPC scheme strictly outperforms the case of no power control\footnote{Throughout this paper, no power control means that all transmitters always uses the same constant power for transmission.}. A fundamental constraint on the discrete power levels, and their selected probabilities are then discovered, which ensures that such designed DPC leads to strictly better performance in terms of the outage probability and mean signal-to-interference ratio (SIR). This constraint is built on the geometric conservation property of a homogeneous PPP, leading to a better outage-free spatial reuse factor, which has a physical meaning of how many transmitters per unit area on average that could simultaneously transmit without outage. Therefore, motivated by the fact that the received signal power heavily depends on the transmission distance, an $N$-layer DPC scheme is proposed for a cluster that is tessellated into $N$-layer annuli, where a suitable discrete power level is chosen from an $N$-tuple power set according to which layer the intended receiver is located at.  To evaluate the throughput performance of this DPC scheme, the metric of transmission capacity (TC) proposed in~\cite{SWXYJGAGDV05,SWJGANJ10} is used after appropriate modification.

Our second contribution is to characterize the outage probability of each layer in a cluster with the proposed $N$-layer power control and then use it to show that the proposed scheme is essentially ``location-dependent'' when it achieves the upper and lower bounds on the maximum contention intensity. This location-dependent characteristic makes the $N$-layer discrete power control have the capability of achieving power saving, interference reduction, and throughput fairness. Since the bounds on the maximum contention intensity are explicitly established, the corresponding TC can also be easily bounded, which indicates how the $N$-layer discrete power control can monotonically increase TC if it is properly devised. Analytical and simulation results both show that the bounds on the achievable outage probability and spatial reuse factor are better than other existing power control schemes.

Our third contribution is outlined as follows. The location-dependent characteristic of the $N$-layer DPC scheme can be generalized to a power control scaling law, i.e., for an intended receiver located at the $i$th layer of a cluster, the transmit power $P_i\in\Theta\left(\eta^{-\frac{\alpha}{2}}_i\right)$ should be used, where $\alpha>2$ is the path loss exponent and $\eta_i$ is the probability of selecting power $P_i$, which usually depends on the area of the $i$th layer. This power control scaling law can not only balance the interference across $N$ different layers, but also reveal how the upper bound of $N$ and the spatial reuse factor change with $\eta_i$. With this power control scaling law, some optimization problems, such as minimizing the sum power over all $P_i$'s or minimizing the mean outage probability over $N$, can be easily formulated. Finally, two examples with different distributions of intended receivers are discussed, whose simulation results show that the proposed $N$-layer DPC can achieve a significantly higher TC than other power control schemes.

\section{System Model and Preliminaries}\label{sec:model}

\subsection{Poisson-Clustered Network Model and Geometric Conservation Property}
In this paper, we consider an infinitely large wireless ad hoc network where transmitters are independently and randomly distributed on the plane $\mathbb{R}^2$, which forms a homogeneous PPP $\Phi$ of intensity $\lambda$ that gives the average number of transmitting nodes per unit area. Each transmitter can have a random number of candidate receivers that are uniformly and randomly distributed in a cluster with a common distribution, independent of the transmitters' spatial distribution. Hence, all the nodes in the network can be viewed to form a Poisson cluster process (PCP) -- A parent (transmitter) node is associated with some daughter (receiver) nodes\footnote{Note that each cluster could contain other transmitters and unintended receivers in addition to its own transmitter and intended receivers.}. The marked transmitter point process $\Phi$ can be expressed as
\begin{equation}
\Phi\defn\{(X_i,P_i,H_i): X_i\in\mathcal{B}_i, P_i,H_i\geq 0, i\in\mathbb{N}\},
\end{equation}
where $X_i$ denotes transmitter $i$ and its location, $P_i$ represents the transmit power of $X_i$, $\mathcal{B}_i$ is the cluster that $X_i$ belongs to, and $H_i$ is the fading channel gain from $X_i$ to its selected receiver $Y_i\in\mathcal{B}_i$. Also, the network is assumed to be interference-limited and operating with a slotted Aloha protocol\footnote{With the slotted Aloha protocol, the interference received by each receiver in the network is merely generated by the transmitting nodes in the current time slot. The interference generated in the previous time slot is not received.}.

A communication link from one node to another in the network experiences path loss and Rayleigh fading. The fading channel power gains of all links are i.i.d. exponential random variables with unit mean and variance. Without loss of generality, transmitter $X_0$ is assumed to be located at the origin and it selects one of the candidate receivers in cluster $\mathcal{B}_0$ for transmission. Thus, we call node $X_0$ the reference transmitter and perform the analysis by conditioning on its receiver (called reference receiver). According to the Slivnyak theorem\cite{Stoyan}\cite{Baccllibook}, the statistics of signal reception seen by the reference receiver is the same as that seen by any receivers of all other transmitter-receiver pairs. The signal-to-interference ratio (SIR) at the reference receiver can be written as
\begin{equation}\label{Eqn:DefnSIR}
\mathrm{SIR}_0 (P_0)=\frac{P_0H_0}{R^{\alpha}I_0},
\end{equation}
where $R$ is the (random) distance from transmitter $X_0$ to its selected receiver $Y_0$, $(\mathtt{distance})^{-\alpha}$ is the pass loss model\footnote{This path-loss model is unreasonable for the near-field nodes with $\|X\|<1$; but we still use it for $\|X\|<1$ since it only makes a negligible effect on our outage probability results \cite{FBBBPM06,SWJGANJ07}.} with path loss exponent $\alpha>2$, and $I_0$ denotes the interference at $Y_0$ given by
$$I_0=\sum_{ X_k \in \Phi \setminus X_0 }{P_k H_{k0} \|X_k - Y_0\|^{-\alpha}},$$
where $\|X_k-Y_0\|$ is the Euclidean distance between interfering transmitter  $X_k$ and $Y_0$, $H_{k0}$ is the fading gain from $X_k$ to $Y_0$, and $P_k$ denotes the transmit power of $X_k$. In order to have a successful signal reception at receiver $Y_0$, the $\text{SIR}$ has to be no less than a predesignated threshold $\beta$; otherwise an outage occurs. Without loss of generality, the outage probability for transmissions using power $P_0$ is thus defined as $\mathbb{P}[\mathrm{SIR}_0(P_0)<\beta]$.

A homogeneous PPP has a nice conservation property, which provides the relationship on how uniform node position scaling changes with the node intensity~\cite{Stoyan}. Here we give the conservation property in the Poisson cluster process (PCP) context with the following lemma.
\begin{lemma}[The Geometric Conservation Property of a PCP] \label{Lam:ConserversionPropertyPCP}
Assume that for each transmitter, the average number of intended receivers in the cluster is $\omega$ and thus all the nodes in the network also form a homogeneous PPP $\Pi$ with intensity $\omega\lambda$. Let $\mathbf{T}: \mathbb{R}^2 \rightarrow \mathbb{R}^2$ be a non-singular transformation matrix in $\mathbb{R}^2$. Then $\mathbf{T}(\Pi)\defn\{\mathbf{T}Z_i:Z_i\in\Pi\}$ is also a homogeneous PPP  with intensity $\omega\lambda/\sqrt{\det(\mathbf{T}^{\emph{\textsf{T}}}\mathbf{T})}$.
\end{lemma}
\begin{IEEEproof}
The void probability of a point process in a bounded Borel set $\mathcal{A}\subset \mathbb{R}^2$ is the probability that $\mathcal{A}$ does not contain any points of the process. Since $\Pi$ is a homogeneous PPP, its void probability is given by
\begin{equation}\label{Eqn:VoidProb}
\mathbb{P}[\Pi(\mathcal{A})=0]=\exp(-\omega\lambda \mu(\mathcal{A})),
\end{equation}
where $\mu(\cdot)$ is a Lebesgue measure in $\mathbb{R}^2$. Since the void probability completely characterizes the statistics of a PPP, we only need to show that the void probability of $\mathbf{T}(\Pi(\mathcal{A}))$ is given by
\begin{equation}
\mathbb{P}[\mathbf{T}(\Pi(\mathcal{A}))=0]=\exp\left(-\omega\lambda/\sqrt{\det(\mathbf{T}^{\textsf{T}}\mathbf{T})}\mu(\mathbf{T}(\mathcal{A}))\right).
\end{equation}

Recall the result from vector calculus that the absolute value of the determinant of a matrix is equal to the volume of the parallelepiped that is spanned by the vectors of the matrix. Therefore, the $2$-dimensional volume of $\mathbf{T}(\mathcal{A})$ is given by $\mu(\mathbf{T}(\mathcal{A}))=\sqrt{\det(\mathbf{T}^{\textsf{T}}\mathbf{T})}\mu(\mathcal{A})$. Suppose $\mathbf{T}(\Pi)$ has intensity $\lambda^{\dag}$ and its void probability within the volume of $\mathbf{T}(\mathcal{A})$ is
\begin{align*}
\mathbb{P}[\mathbf{T}(\Pi(\mathcal{A}))=0]&=\mathbb{P}[\Pi(\mathcal{A})=0]\\
&=\exp\left(-\lambda^{\dag}\sqrt{\det(\mathbf{T}^{\textsf{T}}\mathbf{T})}\mu(\mathcal{A})\right).
\end{align*}
Then by comparing the above equation with \eqref{Eqn:VoidProb}, it follows that $\lambda^{\dag} = \omega\lambda/\sqrt{\det(\mathbf{T}^{\textsf{T}}\mathbf{T})}$.
\end{IEEEproof}
For a special case, if $\mathbf{T}=\sqrt{a}\mathbf{I}_2$ which $\mathbf{I}_2$ a $2\times 2$ identity matrix and constant $a>0$, the intensity of $\mathbf{T}(\Pi)$ changes to $\frac{\omega\lambda}{a}$. Lemma \ref{Lam:ConserversionPropertyPCP} can be used to eliminate the inconsistency in the distribution of interferences induced by multiple transmit power levels adopted in the network, as shown in the following subsection.

\subsection{Why Discrete Power Control?}
As aforementioned, discrete power control is preferable for implementation in practice. There are also two main motivations for adopting discrete power control even from a theoretical point of view. First of all, we show that if a transmitter can control its discrete powers appropriately, its receiver is able to achieve a lower outage probability compared with no power control.
\begin{theorem}\label{Thm:AvgSIRInq}
Consider a special case in the PCP-based network where each cluster contains one transmitter-receiver pair. Each transmitter has $N$ constant power options from the discrete power control set $\mathcal{P}\defn\{P_1, P_2, \cdots, P_N\}$. Suppose each transmitter independently selects its own transmit power and the probability of selecting $P_j\in\mathcal{P}$ is $\eta_j$. The average SIR achieved by transmitters using $N$ discrete powers is strictly greater than that achieved by transmitters using a single constant power if
\begin{equation}\label{Eqn:AvgSIRIneq}
\sum_{j=1}^{N}\eta_j^{\frac{\alpha}{2}}\left(\frac{P_j}{P_i}\right)< \frac{1}{\rho_0},\,\ i\in\{1, 2, \ldots, N\},
\end{equation}
where $\rho_0\defn \mathbb{E}[I_0(1)]\mathbb{E}[I^{-1}_0(1)]\geq 1$ is a function of intensity $\lambda$ and path loss exponent $\alpha$, and $I_0(\nu)\defn \nu (\sum_{X_i\in\Phi\setminus X_0}H_{i0}\|X_i-Y_0\|^{-\alpha})$ denotes the interference at $Y_0$ induced by all interferers in $\Phi$ using transmit power $\nu$. Most importantly, condition \eqref{Eqn:AvgSIRIneq} also ensures that the outage probability achieved by transmitters using $N$ discrete powers is also strictly smaller than that achieved by transmitters using a single constant power.
\end{theorem}
\begin{IEEEproof}
See Appendix  \ref{App:ProofAvgSIRInq}.
\end{IEEEproof}
\begin{remark}
The inequality in \eqref{Eqn:AvgSIRIneq} ensures that discrete power control has a better performance in terms of the average SIR and outage probability than no power control. It can be relaxed to $\sum_{j=1}^{N}\eta_j^{\frac{\alpha}{2}}\left(\frac{P_j}{P_i}\right)< 1$ if we only require a lower outage probability (i.e. no SIR requirement).
\end{remark}
\begin{remark}
The average of the interference $I_0$, $\mathbb{E}[I_0]$, is unbounded since the pass loss model $\|\cdot\|^{-\alpha}$ is not well-defined for very nearby interferers and even explodes at distance zero. To obtain a bounded $\rho_0$, we define $\mathbb{E}[I_0(\nu)]\defn 2\pi \lambda\nu \int_1^{\infty} r^{1-\alpha}\dif r=\frac{2\pi\lambda\nu}{\alpha-2}$, which is obtained by applying the Campbell theorem  \cite{Stoyan} and ignoring the interference contributed by the interferers within the disc with a center at the origin and unit radius.  
\end{remark}

\begin{figure}[t!]
\centering
\includegraphics[width=3.6in,height=2.6in]{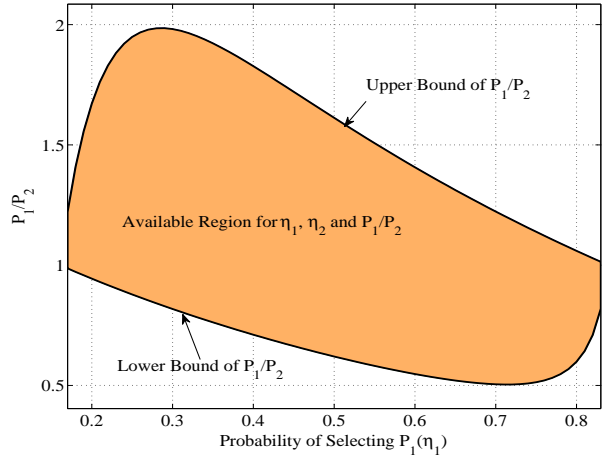}
\caption{The available region of $\frac{P_1}{P_2}$ for $\alpha=3.5$, $\lambda=0.0005$ and $\mathbb{E}[I_0(1)]\mathbb{E}[I_0^{-1}(1)]\approx1.29$. Two discrete powers outperforms a single constant power in terms of the average SIR and outage probability if their ratio is within the colored region.}
\label{fig:RegionDPC}
\end{figure}

\begin{figure}[t!]
\centering
\includegraphics[width=3.6in,height=2.6in]{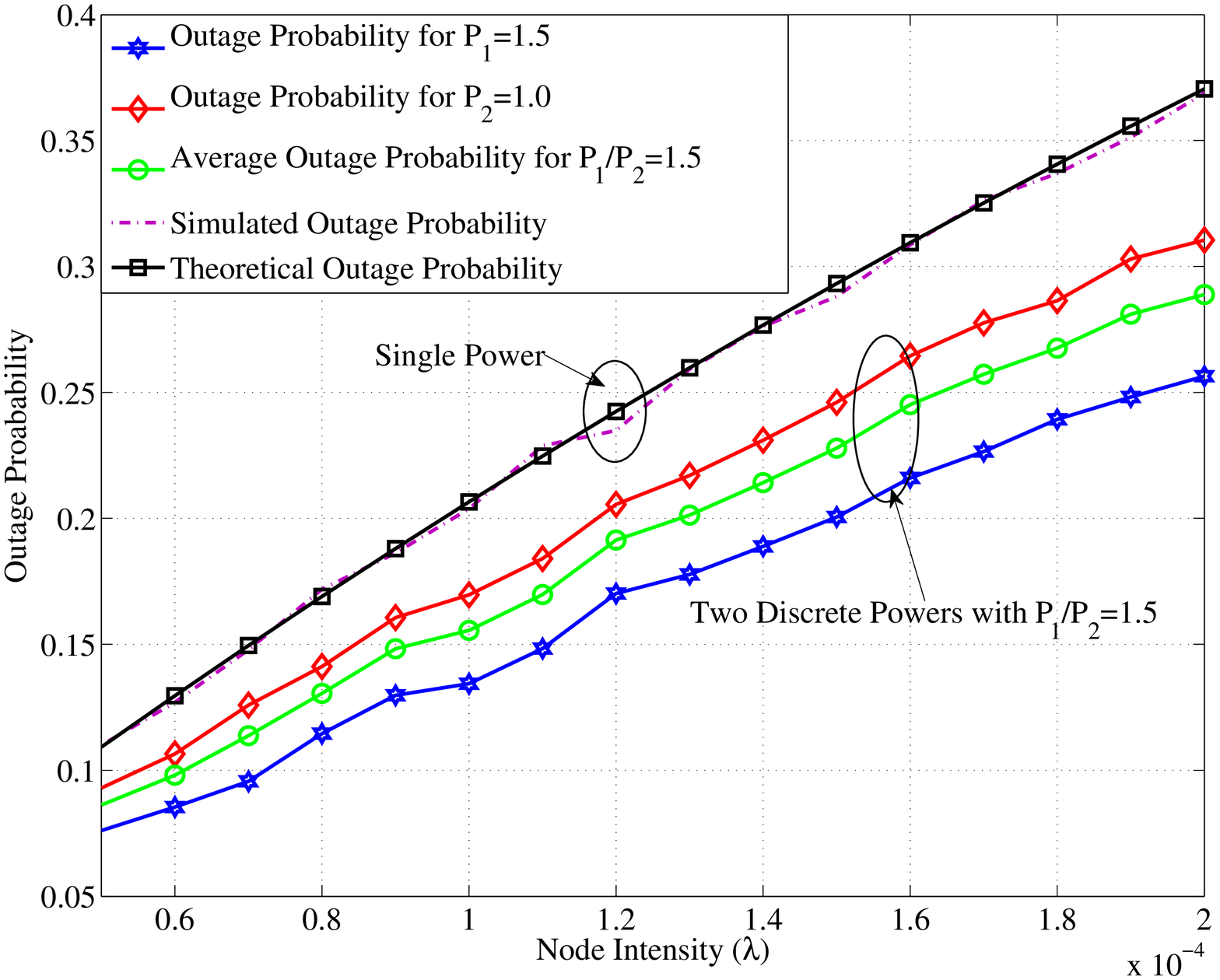}
\caption{The outage probabilities of using two discrete powers and a single power for $\alpha=3.5$, $R=20$m, $\beta=1$, $\eta_1=0.4$ and $\eta_2=0.6$. The ratio of the two discrete powers is $\frac{P_1}{P_2}=1.5$.}
\label{fig:OutProbTwoDisPow}
\end{figure}

Theorem \ref{Thm:AvgSIRInq} indicates that using multiple discrete power level will outperform using no power control if the inequality constraint in \eqref{Eqn:AvgSIRIneq} is satisfied. This is due to the fact that the inequality in \eqref{Eqn:AvgSIRIneq} essentially ensures that the interference generated by multiple transmit powers is not greater than that generated by a single power. In other words, if we use several discrete transmit powers in the network,  a lower outage probability can be attained if those discrete power values and the associated probabilities are properly devised to satisfy \eqref{Eqn:AvgSIRIneq}. For example, if the power control set $\mathcal{P}=\{P_1,P_2\}$ has two tuples, with $P_0$, $P_1$ and $P_2$ distinct, \eqref{Eqn:AvgSIRIneq} can be simplified as
\begin{equation}\label{Eqn:TwoPowRatIneq}
\eta_1^{-\frac{\alpha}{2}}\left(\frac{1}{\rho_0}-\eta_2^{\frac{\alpha}{2}}\right)\geq \frac{P_1}{P_2} \geq \eta_2^{\frac{\alpha}{2}}\left(\frac{1}{\rho_0}-\eta_1^{\frac{\alpha}{2}}\right)^{-1}.
\end{equation}
This result is illustrated in Fig. \ref{fig:RegionDPC} for $\alpha=3.5$ and $\rho_0\approx 1.29$, and the shaded region represents two discrete powers strictly outperform a single power in term of outage. Fig. \ref{fig:OutProbTwoDisPow} illustrates the two outage probabilities and the average outage probability for $R=20$m, $\alpha=3.5$, $\beta=1$, $\eta_1=0.4$, $\eta_2=0.6$, and power ratio $\frac{P_1}{P_2}=1.5$ satisfying \eqref{Eqn:TwoPowRatIneq} where the two outage probabilities and the average outage probability are $\mathbb{P}[\mathrm{SIR}_0(P_1)<\beta]$, $\mathbb{P}[\mathrm{SIR}_0(P_2)<\beta]$ and $\eta_1\mathbb{P}[\mathrm{SIR}_0(P_1)<\beta]+\eta_2\mathbb{P}[\mathrm{SIR}_0(P_2)<\beta]$, respectively. Note that the simulation result for the single power case does not depend what constant power is used since the SIR in (2) does not depend on transmit power in the no power control scheme. As we see, all the outage probabilities with two discrete powers are (much) lower than that with a single power. Moreover, the inequality in \eqref{Eqn:AvgSIRIneq} makes the average SIR with DPC higher than the average SIR without power control, which results in a higher channel capacity bound on average.

Another interesting observation that can be drawn from \eqref{Eqn:AvgSIRIneq} is that it reveals a simple method to design those discrete power values. For example, we can consider $P_i\in\Theta\left(\eta_i^{-\frac{\alpha}{2}}\right)$, which results in  $\min_i\{P_i\}\in\Omega(N)$, that is, $\max_i\{\eta_i\}\in\Omega\left( N^{-\frac{2}{\alpha}}\right)$ \footnote{
Throughout this paper, we slightly relax standard asymptotic notations to denote the scaling results in this paper:  $O(\cdot)$, $\Omega(\cdot)$ and $\Theta(\cdot)$ correspond to (asymptotic) upper, lower, and tight
bounds, respectively. For instance, given two real-valued functions $f(x)$ and $g(x)$, we use $f(x)\in\Theta(g(x))$ to mean that there exist two positive constants $c_1$ and $c_2$ such that $c_1g(x)\leq f(x)\leq c_2 g(x)$ for all $x\in\mathbb{R}$, i.e., $x$ does not have to go to infinitely large or small to make $c_1g(x)\leq f(x)\leq c_2 g(x)$ to hold.}. Thus, the minimum required power can be determined by $N$ and $\lambda$, and we are able to know the minimum number of discrete powers needed once the node intensity and the power $\min_i\{P_i\}$ are known. Usually, selecting transmit power depends on the channel gain condition such that the probabilities $\{\eta_i\}$ are related to some \textit{uncontrollable} network parameters such as the distributions of channel fading and node locations. That implies that the selection of discrete power control can be specified in terms of certain network parameters.

From a \textit{spatial reuse} point of view, we can also explain why using discrete power control can do better. Since the outage probability can be written as $\mathbb{P}\left[(P_0H_0/\beta I_0)^{\frac{1}{\alpha}}<R\right]$, there is no outage once the transmission distance is less than or equal to $(P_0H_0/\beta I_0)^{\frac{1}{\alpha}}$ that is called the\textit{ maximum transmission distance without outage}. Motivated by the similar concept of spatial reuse defined in \cite{FBBBPM06} and the maximum transmission distance without outage, we define the outage-free spatial reuse factor as follows.
\begin{definition}[Outage-Free Spatial Reuse Factor]
The (outage-free) spatial reuse factor $\delta_0$ for transmitter $X_0$ with power $P_0$ is defined by
\begin{equation}\label{Eqn:DefnSpatialReuse}
\delta_0\defn \dfrac{\mathbb{E}\left[\pi (P_0H_0/\beta I_0)^{\frac{2}{\alpha}}\lambda\right]}{\mathbb{E}[\pi D^2_0 \lambda]}=\pi\lambda\,\mathbb{E}\left[\left(\frac{P_0H_0}{I_0\beta}\right)^{\frac{2}{\alpha}}\right],
\end{equation}
where $D_0$ is the nearest distance between two transmitters and its pdf is $f_{D_0}(x)=2\pi\lambda x e^{-\pi\lambda x^2}$ and $\mathbb{E}[D_0^2]=\frac{1}{\pi\lambda}$.
\end{definition}
\noindent According to \eqref{Eqn:DefnSpatialReuse}, the physical meaning of the spatial reuse factor can be interpreted as the average number of transmitting nodes that can coexist in the defined maximum outage-free (circular) transmission area. The larger the spatial reuse factor is, the higher the effective network throughput per unit area is. Note that for the case of no power control, $\delta_0$ becomes
\begin{equation}
\delta_0^{\textrm{np}}=\pi\lambda \Gamma\left(1+\frac{2}{\alpha}\right)\beta^{-\frac{2}{\alpha}}\mathbb{E}\left[I_0^{-\frac{2}{\alpha}}(1)\right],
\end{equation}
where $\Gamma(x)=\int_{0}^{\infty}t^{x-1}e^{-t}\dif t$ is the Gamma function and $\mathbb{E}\left[I^{-\frac{2}{\alpha}}_0(1)\right]$ is lower-bounded by $(\mathbb{E}[I_0(1)])^{-\frac{2}{\alpha}}=\left(\frac{\alpha-2}{2\pi\lambda}\right)^{\frac{2}{\alpha}}$. That means $\mathbb{E}\left[I^{-\frac{2}{\alpha}}_0(1)\right]\in\Omega(\lambda^{-\frac{2}{\alpha}})$ and thus the spatial reuse factor for no power control is $\delta^{\textrm{np}}_0=\frac{\delta_0}{P_0} \in\Omega(\lambda^{1-\frac{2}{\alpha}})$. Thus $\delta^{\textrm{np}}_0$ increases when $\lambda$ increases, which means the shrinking speed of the average outage-free area is slower than that of the average area without any transmitters.

In order to increase the spatial reuse factor, we can appropriately control transmit power. The following lemma will show how the spatial reuse factor under a DPC can be increased.
\begin{lemma}\label{Lem:SpatialReuse}
In a Poisson-distributed wireless network with transmitter intensity $\lambda$, each transmitter independently selects power $P_i$ from power set $\mathcal{P}=\{P_1, P_2, \ldots, P_N\}$ with probability $\eta_i$. If all discrete powers and their corresponding selected probabilities satisfy \eqref{Eqn:AvgSIRIneq}, the spatial reuse factor induced by transmitters with discrete power $P_i$ is $\delta^{\textrm{dp}}_{0_i}\defn \mathbb{E}[(H_0/\beta(I_0/P_i))^{\frac{2}{\alpha}}]/\mathbb{E}[D_0^2]$ that is greater than $\delta^{\textrm{np}}_0$.  The average spatial reuse factor with discrete power control $\mathcal{P}$ is defined as
\begin{equation}\label{Eqn:AvgSpaReuFac}
\delta^{\textrm{dp}}_0\defn \sum_{i=1}^{N}\eta_i\delta^{\textrm{dp}}_{0_i},
\end{equation}
and thus $\delta^{\textrm{dp}}_0>\delta_0^{\textrm{np}}$ since $\delta^{\textrm{dp}}_{0_i}>\delta_0^{\textrm{np}}$ for all $i$.
\end{lemma}
\begin{IEEEproof}
First consider the case of no power control and the maximum transmission distance without outage in this case, which is $(H_0/\beta I_0(1))^{\frac{1}{\alpha}}$. By definition, the spatial reuse factor $\delta^{\textrm{np}}_0$ is given by
\begin{equation}
\delta^{\textrm{np}}_0=\lambda\pi\Gamma\left(1+\frac{2}{\alpha}\right)\beta^{-\frac{2}{\alpha}}\mathbb{E}\left[I_0^{-\frac{2}{\alpha}}(1)\right].
\end{equation}
Now consider that transmitter $X_j\in\Phi$ uses discrete power $P_j\in\mathcal{P}$ with probability $\eta_j$ and thus the receiver $Y_0$ of transmitter $X_0$ using power $P_i$ experiences the following interference normalized by $P_i$
\begin{align*}
\frac{I_0}{P_i}&= \sum_{j=1}^{N}\frac{P_j}{P_i}\sum_{X_k\in\Phi_j}H_{k0}\|X_k\|^{-\alpha}\stackrel{d}{=} \sum_{j=1}^{N}\sum_{X_k\in\Phi'_j}H_{k0}\|X_k\|^{-\alpha}\\
&\stackrel{d}{=}\sum_{j=1}^{N}\eta_j^{\frac{\alpha}{2}}\left(\frac{P_j}{P_i}\right)\sum_{X_m\in\hat{\Phi}_j}H_{m0}\|X_m\|^{-\alpha},
\end{align*}
where $\Phi'_j$ is a PPP of intensity $\lambda\eta_j(P_j/P_i)^{\frac{2}{\alpha}}$ and $\hat{\Phi}_j$ is a PPP of intensity $\lambda$. Whereas the spatial reuse factor $\delta_{0_i}$ induced by $X_0$ with power $P_i$ can be equivalently  defined as
\begin{align*}
\delta^{\textrm{dp}}_{0_i}&\defn \mathbb{E}\left[\left(\frac{\beta \sum_{j=1}^{N}\sum_{X_k\in\Phi'_j} H_{k0}\|X_k\|^{-\alpha}}{H_0}\right)^{-\frac{2}{\alpha}}\right]/\left(\mathbb{E}[D^2_0]\right)\\
&\geq\delta^{\textrm{np}}_0\left[\sum_{j=1}^{N}\eta_j^{\frac{\alpha}{2}}\left(\frac{P_j}{P_i}\right)\right]^{-\frac{2}{\alpha}}\frac{(\mathbb{E}[I_0(1)])^{-\frac{2}{\alpha}}}{\mathbb{E}\left[I^{-\frac{2}{\alpha}}_0(1)\right]}
\end{align*}
\begin{align}
&= \delta^{\textrm{np}}_0\left[\rho_0\sum_{j=1}^{N}\eta_j^{\frac{\alpha}{2}}\left(\frac{P_j}{P_i}\right)\right]^{-\frac{2}{\alpha}}\frac{(\mathbb{E}[I^{-1}_0(1)])^{\frac{2}{\alpha}}}{\mathbb{E}\left[I^{-\frac{2}{\alpha}}_0(1)\right]}.\label{Eqn:LowBouSpaReuPowi}
\end{align}
Since $(\mathbb{E}[I^{-1}_0(1)])^{\frac{2}{\alpha}}\geq \mathbb{E}\left[I^{-\frac{2}{\alpha}}_0(1)\right]$, we can make sure $\delta^{\textrm{dp}}_{0_i}> \delta_0^{\textrm{np}}$ whenever $\rho_0\sum_{j=1}^{N}\eta_j^{\frac{\alpha}{2}}\left(\frac{P_j}{P_i}\right)<1$.
Thus it follows that $\delta^{\textrm{dp}}_{0_i}>\delta^{\textrm{np}}_0$ if the condition in \eqref{Eqn:AvgSIRIneq} is satisfied. Substituting the above result of $\delta^{\textrm{dp}}_{0_i}$ into the definition of $\delta_0^{\textrm{dp}}$ leads to \eqref{Eqn:AvgSpaReuFac}.
\end{IEEEproof}

The inequality in \eqref{Eqn:AvgSIRIneq} for spatial reuse ensures that the effect of discrete powers and their corresponding probabilities is able to geometrically lessen the scaling of the transmitter intensity. This point can be further illustrated by taking a closer look at  the average spatial reuse factor $\delta^{\textrm{dp}}_0$ in \eqref{Eqn:LowBouSpaReuPowi} via the following form:
$$\delta^{\textrm{dp}}_0> \lambda\pi\Gamma\left(1+\frac{2}{\alpha}\right)\beta^{-\frac{2}{\alpha}}\mathbb{E}\left[\tilde{I}_0^{-\frac{2}{\alpha}}(1)\right],$$
where $\tilde{I}_0(1)$ is the interference generated by a transmitter PPP with unit constant transmit power and intensity $\lambda \left[\rho_0\sum_{j=1}^{N}\eta_j^{\frac{\alpha}{2}}\left(\frac{P_j}{P_i}\right)\right]^{\frac{2}{\alpha}}\mathbb{E}\left[I^{-\frac{2}{\alpha}}_0(1)\right]/(\mathbb{E}[I^{-1}_0(1)])^{\frac{2}{\alpha}}$, which is smaller than $\lambda$. Hence, the average number of coexisting transmitters without outage per unit area is increased. 

\begin{figure}[t!]
\centering
\includegraphics[width=3.6in,height=2.75in]{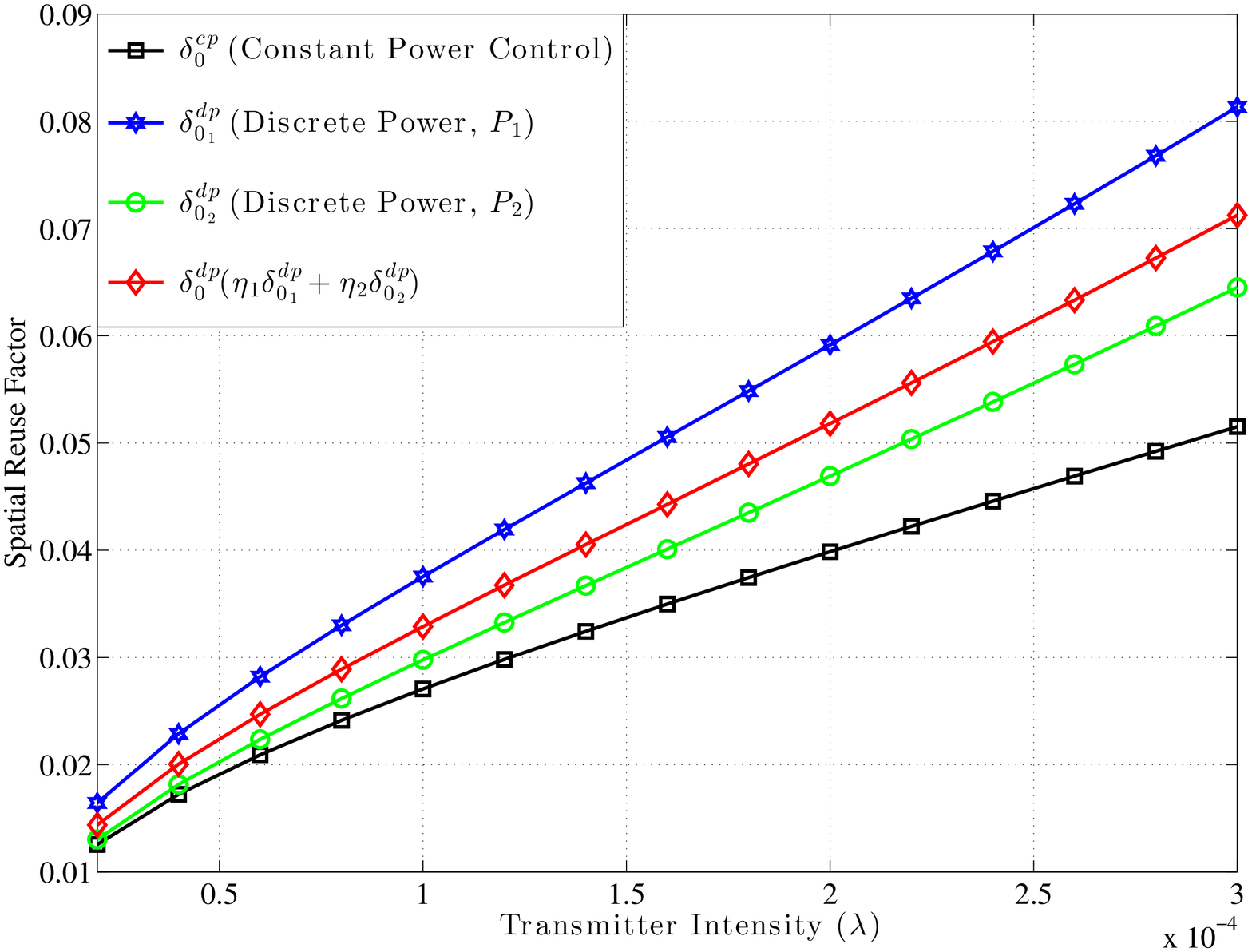}
\caption{The spatial reuse factors for discrete power control and no power control schemes. The network parameters for simulation are $\alpha=3.5$, $\frac{P_1}{P_2}=1.5$, $\eta_1=0.4$ and $\eta_2=0.6$.}
\label{fig:SpaResFac}
\end{figure}

Although the spatial reuse factor characterizes how space is effectively used for simultaneous successful transmissions, it fails to characterize the temporal transmission efficiency of a communication link. Reducing the outage probability certainly increases the temporal transmission efficiency since it results in fewer retransmission behaviors. Surprisingly, here we see that the condition in \eqref{Eqn:AvgSIRIneq} is able to guarantee a better spatial reuse factor as well as a lower outage probability.  That is, both spatial and temporal transmission efficiencies can be enhanced if all discrete powers and their corresponding probabilities satisfy  \eqref{Eqn:AvgSIRIneq}. Therefore,  \eqref{Eqn:AvgSIRIneq} is the fundamental requirement to ensure that discrete power control is strictly superior to no power control. The simulation results of how the spatial reuse factors with two discrete powers are superior to the spatial reuse factor with a single power are shown in Fig. \ref{fig:SpaResFac} by assuming $\alpha=3.5$, $\frac{P_1}{P_2}=1.5$, $\eta_1=0.4$ and $\eta_2=0.6$. Finally, \eqref{Eqn:AvgSIRIneq} also motivates us a simple discrete power design approach. For example, we can adopt $P_i\in\Theta\left(\eta_i^{-\frac{\alpha}{2}}\right)$ as the power design in the case of reducing outage probability, and then   \eqref{Eqn:AvgSIRIneq} gives $\min_i\{P_i\}\in \Omega(N^{\frac{\alpha}{2}})$, i.e., $\max_i\{\eta_i\}\in O\left(\frac{1}{N}\right)$. The required $N$ and discrete powers $\{P_i\}$ can be properly chosen once the probabilities $\{\eta_i\}$ related to network parameters are determined. In Section \ref{Sec:NlayerDPC}, we will show that the DPC scaling law $P_i\in\Theta\left(\eta_i^{-\frac{2}{\alpha}}\right)$ is a general expression for increasing TC with $N$-layer DPC.

\section{$N$-Layer Discrete Power Control}\label{Sec:NlayerDPC}
Since signal power decays heavily over the transmission distance, it is nature for us to consider an $N$-layer DPC scheme that is devised based on the transmission distance to the intended receiver in a cluster, i.e., we consider a cluster tessellated into $N$-layer annuli and each time a transmitter selects one receiver at a certain layer of the cluster for service. If the selected receiver is at the $i$th layer, power $P_i$ is used for transmission at transmitter $X_0$, where the outage probability at receiver $Y_0$  is given by
\begin{equation}\label{eq:def+OP}
q_i \defn \mathbb{P}[\mathrm{SIR}_0(P_i)<\beta],\quad i\in[1,\ldots,N].
\end{equation}
This outage probability for a receiver located at the $i$th layer can be used to define the following transmission capacity in the $N$-layer DPC context.
\begin{definition}[Transmission Capacity with $N$-layer DPC] The transmission capacity for the $N$-layer DPC scheme is defined by
\begin{equation}\label{eq:DefnTC}
C^{\textrm{dp}}_{\epsilon} \defn \gamma\, \lambda^{\textrm{dp}}_{\epsilon}\sum_{i=1}^N \eta_i \left[1-q_i(\lambda^{\textrm{dp}}_{\epsilon})\right],
\end{equation}
where $\eta_i$ denotes the fraction of intended receivers being served in the $i$th layer, $\gamma$ is the transmission rate per unit bandwidth of each communication link, and $\lambda^{\textrm{dp}}_{\epsilon}$, called the maximum contention intensity, is given by
\begin{equation}
\lambda^{\textrm{dp}}_{\epsilon}\defn \sup\left\{\lambda : \max_{i\in\{1, 2, \cdots, N\}} q_i(\lambda)\leq \epsilon\right\},
\end{equation}
where $\epsilon$ denotes the upper bound on the outage probability and usually it is a small number.
\end{definition}
\noindent Note that the transmission capacity defined in \eqref{eq:DefnTC} represents the area spectrum efficiency of the $N$-layer DPC, which is different from  and actually a generalized form of the transmission capacity originally proposed in \cite{SWJGANJ07} for the point-to-point communication scenario. It degrades to the original one when each cluster only contains one intended receiver and there is no power control.

Suppose that the distance $R$ from a transmitter to its intended receiver in a cluster is a random variable whose probability density function (pdf) and cumulative density function (cdf) are denoted by  $f_{R}(r)$ and $F_R(r)$, respectively. Our $N$-layer DPC scheme is to use a transmit power based on which layer the selected intended receiver is located at. Let the maximum transmission distance in a cluster $\mathcal{B}$ be quantized into $N$ intervals, i.e., $\{\mathcal{L}_i, i=1,2,\ldots,  N\}$,  where $\mathcal{L}_i$ is the $i$th interval with $\bigcup_{i=1}^N \mathcal{L}_i \subseteq \mathcal{B}$, and receivers are at layer $i$ if the distances from their transmitter are in interval $\mathcal{L}_i$. A transmitter transmits to its layer-$i$ receivers with the $i$th transmit power chosen from power set $\mathcal{P}=\{P_1, P_2, \ldots, P_N\}$. Then the average outage probability of the layer-$i$ receivers is given in the following theorem.
\begin{theorem}\label{Thm:OutProbLayi}
The average outage probability at the layer-$i$ receivers is given by
\begin{eqnarray}
\label{Eqn:OutProbLayer_i}
q_i= 1-\mathbb{E}\left[e^{-\lambda T_i\beta^{\frac{2}{\alpha}}R^2}\bigg|R\in \mathcal{L}_i\right],
\end{eqnarray}
where $T_i  = \kappa_{\alpha}\sum_{j=1}^{N}{\eta_j \left(\frac{P_j}{P_i}\right)^{\frac{2}{\alpha}}}$ and $\eta_i =\mathbb{P}[R\in\mathcal{L}_i]$.
\end{theorem}
\begin{IEEEproof}
See Appendix \ref{App:ProofOutProbLayi}.
\end{IEEEproof}

For a general distance distribution, the result in \eqref{Eqn:OutProbLayer_i} cannot be further reduced to a closed-form expression. For special cases, consider the one that receivers are uniformly distributed around their transmitter in a circular cluster of radius $s$. In this case, the cdf and pdf of distance $R$ become
\begin{eqnarray*}
F_R(r) = \frac{r^2}{s^{2}}  \quad\text{and}\quad f_R(r) = \frac{2r}{s^2}, \text{ respectively}.
\end{eqnarray*}
Substituting the above $F_R(r)$ and $f_R(r)$ into \eqref{Eqn:OutProbLayer_i} and applying $\eta_i= \mathbb{P}[R\in\mathcal{L}_i]=\frac{1}{s^2}[(\sup(\mathcal{L}_i))^2-(\inf(\mathcal{L}_i))^2]$, the average outage probability for the layer-$i$ receivers becomes
\begin{eqnarray}\label{Eqn:OutProbUniDisR}
q_i  = 1- \frac{ e^{-\lambda T_i\beta^{\frac{2}{\alpha}}(\inf(\mathcal{L}_i))^2 }  - e^{-\lambda T_i\beta^{\frac{2}{\alpha}}(\sup(\mathcal{L}_i))^2}}{ \lambda T_i\beta^{\frac{2}{\alpha}}[(\sup(\mathcal{L}_i))^2-(\inf(\mathcal{L}_i))^2]},
\end{eqnarray}
which can be approximated by
\begin{equation}\label{Eqn:OutProbUniDisRApp}
q_i \approx \frac{1}{2}\lambda T_i\beta^{\frac{2}{\alpha}}\left[(\sup(\mathcal{L}_i))^2+(\inf(\mathcal{L}_i))^2\right]
\end{equation}
when $\lambda$ is small. In addition, an important implication that can be grasped from Theorem \ref{Thm:OutProbLayi} is that the optimal power control that maximizes  $\lambda_{\epsilon}^{\mathrm{dp}}$ depends on the distribution of the receiver distance $R$. The following theorem shows that there exists a (location-dependent) $N$-layer DPC scheme that could achieve (tight) upper and lower bounds on $\lambda_{\epsilon}^{\textrm{dp}}$.
\begin{theorem}\label{Thm:PowerAlloBoundLambda}
If all intended receivers in each cluster of radius $s$ are uniformly distributed and the power ratio of $P_j$ to $P_i$ is set as
\begin{eqnarray}\label{eq:cont+lower+power}
\frac{P_j}{P_i} = \left[\frac{(\sup(\mathcal{L}_j))^2+(\inf(\mathcal{L}_j))^2}{(\sup(\mathcal{L}_i))^2+(\inf(\mathcal{L}_i))^2}\right]^{\frac{\alpha}{2}},
\end{eqnarray}
the lower bound on $\lambda^{\textrm{dp}}_{\epsilon}$ given as
\begin{equation}\label{eq:cont+lambda+lb}
\underline{\lambda}_{\epsilon}^{\textrm{dp}}=\frac{2\epsilon s^2}{\kappa_{\alpha}\beta^{\frac{2}{\alpha}}\sum_{j=1}^{N}[(\sup(\mathcal{L}_j))^4-(\inf(\mathcal{L}_j))^4]}
\end{equation}
could be achieved. If the following power ratio constraints
\begin{equation}\label{eq:cont+upper+power}
\frac{P_j}{P_i} = \left[\frac{\inf(\mathcal{L}_j)}{\inf(\mathcal{L}_i)}\right]^{\alpha}
\end{equation}
are satisfied for all $i\neq j$, the upper bound on $\lambda^{\textrm{dp}}_{\epsilon}$ given as
\begin{equation}\label{eq:cont+lambda+ub}
\overline{\lambda}_{\epsilon}^{\textrm{dp}}=\frac{\epsilon s^2}{(1-\epsilon)\kappa_{\alpha} \beta^{\frac{2}{\alpha}} \sum_{j=1}^{N}[(\sup(\mathcal{L}_j)\inf(\mathcal{L}_j))^2-(\inf(\mathcal{L}_j))^4] }
\end{equation}
could be achieved.
\end{theorem}
\begin{IEEEproof}
According to the proof of Theorem \ref{Thm:OutProbLayi}, the outage probability associated with  layer $i$ is given by
\begin{eqnarray}
q_i = \frac{1}{\eta_i s^2}\int_{\mathcal{L}_i}\left(1-e^{- \lambda T_i\beta^{\frac{2}{\alpha}}r^2 }\right) \dif r^2
\end{eqnarray}
By utilizing the fact that $\frac{x}{1+x}\leq 1-e^{-x}\leq x$ for $x\geq 0$, the outage probability $q_i$ is upper-bounded by
\begin{align}
q_i &\leq \frac{2\lambda T_i\beta^{\frac{2}{\alpha}}}{\eta_is^2}\int_{\mathcal{L}_i} r^3 \dif r\nonumber\\ 
&= \frac{\lambda T_i\beta^{\frac{2}{\alpha}}}{2 }\left[(\sup(\mathcal{L}_i))^2+(\inf(\mathcal{L}_i))^2\right]\defn \overline{q}_i. \label{Eqn:UppBound_q_i}
\end{align}
Since $q_i$ is a continuous and monotonic increasing function of the intensity $\lambda$, the maximum contention intensity $\lambda^{\mathrm{dp}}_{\epsilon}$ that makes $q_i$ equal to $\epsilon$  must exist. As a result, the intensity, as defined in the following
\begin{eqnarray}
\underline{\lambda}^{\textrm{dp}}_i \defn \sup \{\lambda: \overline{q}_{i}(\lambda)= \epsilon, 1 \leq i \leq N  \},
\end{eqnarray}
satisfies $\overline{q}_i  = \epsilon$, which  is indeed a lower bound on the maximum contention intensity. Hence, it follows that
\begin{align}\label{Eqn:LowBoundIntensityLayer_i}
\underline{\lambda}^{\textrm{dp}}_i= \frac{2 \epsilon }{\beta^{\frac{2}{\alpha}}T_i\left[(\sup(\mathcal{L}_i))^2+(\inf(\mathcal{L}_i))^2\right]},
\end{align}
 for all $i \in \left[1, 2, \ldots, N\right]$.

Next we  explain that the maximum $\lambda$ satisfying $\max_{i}\overline{q}_i = \epsilon $ is attained when all $\overline{q}_i$ are equal. Define $\underline{\lambda}^{\textrm{dp}}_{\epsilon}\defn \min_{i}\{\underline{\lambda}_i^{\textrm{dp}} \}$, which is the intensity that makes the outage probability at each layer less than or equal to $\epsilon$. Since   $\lambda_{\epsilon}^{\mathrm{dp}} \geq \underline{\lambda}_i^{\textrm{dp}}$  for all $i \in [1, 2, \ldots,N]$,  it follows that  $\lambda_{\epsilon}^{\mathrm{dp}}\geq\underline{\lambda}^{\textrm{dp}}_{\epsilon}$ by definition. Thus $ \underline{\lambda}_{\epsilon}^{\textrm{dp}} $ can be maximized up to $\lambda^{\mathrm{dp}}_{\epsilon}$ if all $\overline{q}_i$'s are equal to $\epsilon$, i.e., $\overline{q}_1= \overline{q}_2 = \cdots = \overline{q}_N=\epsilon$.
This equality constraint results in the following power ratio condition:
\begin{eqnarray}
\frac{P_j}{P_i} = \left[\frac{(\sup(\mathcal{L}_j))^2+(\inf(\mathcal{L}_j))^2}{(\sup(\mathcal{L}_i))^2+(\inf(\mathcal{L}_i))^2}\right]^{\frac{\alpha}{2}}.
\end{eqnarray}
By substituting the above power ratio into \eqref{Eqn:LowBoundIntensityLayer_i}, $\underline{\lambda}_{\epsilon}^{\mathrm{dp}}$ given in \eqref{eq:cont+lambda+lb} is obtained.

To obtain an upper bound on the maximum contention intensity, we use $1-e^{-x}\geq\frac{x}{1+x}$ to find the lower bound on the outage probability at layer $i$ as
\begin{align}
q_i &\geq \frac{1}{\eta_is^2}\int_{\mathcal{L}_i}\frac{\lambda T_i\beta^{\frac{2}{\alpha}}r^2}{1+\lambda T_i\beta^{\frac{2}{\alpha}}r^2}\dif r^2\nonumber \\
&= 1-\frac{1}{\eta_i s^2}\int_{\mathcal{L}_i}\frac{1}{1+\lambda T_i\beta^{\frac{2}{\alpha}}r^2}\dif r^2
\end{align}
\begin{align} 
&= 1- \frac{1}{\eta_i s^2 \lambda T_i\beta^{\frac{2}{\alpha}}}\ln\left[\frac{1+\lambda \beta^{\frac{2}{\alpha}}T_i(\sup(\mathcal{L}_i))^2}{1+\lambda \beta^{\frac{2}{\alpha}}T_i(\inf(\mathcal{L}_i))^2}\right]\nonumber\\
&\stackrel{(c)}{\geq} 1-\frac{1}{\eta_i s^2}\left[\frac{(\sup(\mathcal{L}_i))^2-(\inf(\mathcal{L}_i))^2}{1+\lambda \beta^{\frac{2}{\alpha}}T_i(\inf(\mathcal{L}_i))^2}\right] \nonumber\\
&= \frac{\lambda \beta^{\frac{2}{\alpha}}T_i(\inf(\mathcal{L}_i))^2}{1+\lambda \beta^{\frac{2}{\alpha}}T_i(\inf(\mathcal{L}_i))^2}\defn \underline{q}_i, \label{Eqn:LowBound_qi}
\end{align}
where $(c)$ follows from $\ln\left(\frac{1+x}{1+y}\right)\leq \frac{x-y}{1+y}$ for $x>y>0$. Thus $\lambda$ that satisfies $\underline{q}_i =\epsilon$ provides an upper bound on $\lambda_{\epsilon}^{\mathrm{dp}}$. That means
\begin{eqnarray}
\overline{\lambda}^{\textrm{dp}}_i = \sup\{\lambda : \underline{q}_i(\lambda)= \epsilon\}\geq \lambda_{\epsilon}^{\mathrm{dp}},\,\,i\in\{1, 2, \ldots, N\}.
\end{eqnarray}
Similarly, we can argue that $\overline{\lambda}_{\epsilon}^{\mathrm{dp}}=\max_i \{\overline{\lambda}_i^{\mathrm{dp}}\}$ is maximized provided that all $ \underline{q}_i $'s are equal to $\epsilon$. This leads to the power ratio in \eqref{eq:cont+upper+power}. By substituting \eqref{eq:cont+upper+power} into $\underline{q}_i = \epsilon$, $\overline{\lambda}_{\epsilon}^{\mathrm{dp}}$ can be characterized in \eqref{eq:cont+lambda+ub}.
\end{IEEEproof}

Since $\epsilon$ and the node intensity are fairly small for most of practical situations, the upper bound in \eqref{Eqn:UppBound_q_i} and lower bound in \eqref{Eqn:LowBound_qi} on $q_i$ are very tight for all $i$'s since they both approach to $\lambda\beta^{\frac{2}{\alpha}} T_i(\inf(\mathcal{L}_i))^2$ as $\lambda$ is very small, and thus the bounds in \eqref{eq:cont+lambda+lb} and \eqref{eq:cont+lambda+ub} are pretty tight as well. Hence, the power ratios given in \eqref{eq:cont+lower+power} and \eqref{eq:cont+upper+power} could be said to nearly achieve $\lambda^{\mathrm{dp}}_{\epsilon}$ for a given small $\epsilon$ since they achieve the tight bounds on $\lambda^{\mathrm{dp}}_{\epsilon}$. Moreover, those power ratios could achieve network-wise throughput fairness since they have the effect on balancing the outage probabilities for all layers such that the average throughput of  receivers in different layers are almost balanced to the same value. In other words, the throughput degradation problem  between remote and nearby receivers hardly exists.

\begin{figure}[!t]
\centering
\includegraphics[width=3.6in,height=2.75in]{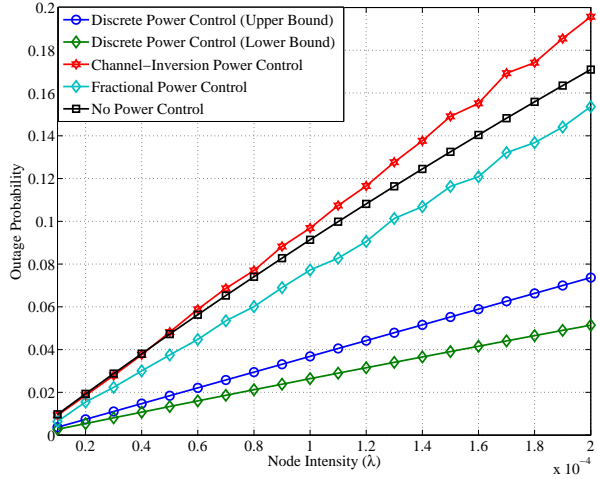}
\caption{The outage probabilities for different power control schemes. The network parameters for simulation are: $\alpha =3.5$, $\beta =1$. The transmit power for each transmitter $X_i$ using fractional power control is $1/\sqrt{H_i}$, while each transmitter $X_i$ using channel-inversion power control has transmit power $1/H_i$.  The transmission distance for fractional power control and channel-inversion power control is a random variable uniformly distributed in [1m, 20m] while the transmission distance 20m is quantized into $N=5$ layers for discrete power control.}
\label{Fig:OutProbUppLowBound}
\end{figure}

If no power control is used, the average outage probability at the $i$th layer becomes
\begin{align*}
\label{eq:cont+npc+qi}
q_i^{\textrm{np}}(\lambda) = 1-\frac{1}{\eta_i}\int_{\mathcal{L}_i}{ \left(\exp\left\{-\lambda \beta^{\frac{2}{\alpha}}\kappa_{\alpha}r^2\right\}\right) \frac{2r}{s^2} \dif r}
\end{align*}
\begin{align}
=& 1-\frac{1}{\lambda\beta^{\frac{2}{\alpha}}\kappa_{\alpha}\eta_i s^2}\exp\left\{-\lambda \beta^{\frac{2}{\alpha}}\kappa_{\alpha}(\inf(\mathcal{L}_i))^2\right\}\cdot\nonumber\\
&\left(1-\exp\left\{-\lambda \beta^{\frac{2}{\alpha}}\kappa_{\alpha} s^2\eta^2_i\right\}\right)
\end{align}
when the intended receivers are uniformly distributed in a cluster. Then \eqref{eq:cont+npc+qi} is lower-bounded as
\begin{align}
q_i^{\textrm{np}}\geq & \frac{\kappa_{\alpha}\beta^{\frac{2}{\alpha}}\lambda \left(\inf(\mathcal{L}_i)\right)^2}{1+\kappa_{\alpha}\beta^{\frac{2}{\alpha}}\lambda \left(\inf(\mathcal{L}_i)\right)^2}\nonumber\\
=&\frac{T_i\beta^{\frac{2}{\alpha}}\lambda \left(\inf(\mathcal{L}_i)\right)^2}{\sum_{j=1}^{N}\eta_j\left(\frac{P_j}{P_i}\right)^{\frac{2}{\alpha}}+T_i\beta^{\frac{2}{\alpha}}\lambda \left(\inf(\mathcal{L}_i)\right)^2}.
\end{align}
Recall that $\frac{T_i}{\kappa_{\alpha}}=\sum_{j=1}^{N}\eta_j\left(\frac{P_j}{P_i}\right)^{\frac{2}{\alpha}}$ and this term is due to discrete power control. Therefore, if we let $\sum_{j=1}^{N}\eta_j\left(\frac{P_j}{P_i}\right)^{\frac{2}{\alpha}}<\frac{1}{\rho_0}$ for all $i\in\{1, 2, \ldots, N\}$, then condition in \eqref{Eqn:AvgSIRIneq} is automatically satisfied and the lower bound $\underline{q}_i$ in the proof of Theorem \ref{Thm:PowerAlloBoundLambda} is smaller than the lower bound on $q_i^{\textrm{np}}$ above. The upper bound $\underline{q}_i$  in the proof of Theorem \ref{Thm:PowerAlloBoundLambda} can be shown to be smaller than  the lower bound on $q_i^{\textrm{np}}$ for most of the practical cases (i.e., $N\geq 2$ and small $\lambda$). So the outage probability performance of the DPC scheme in Theorem \ref{Thm:PowerAlloBoundLambda} is better than that of  no  power control such that a larger transmission capacity could be achieved by the discrete power control. Fig. \ref{Fig:OutProbUppLowBound} shows the simulation results of the outage probabilities for different power control schemes. As can be seen, the bounds corresponding to discrete power control are actually fairly tight  when $\lambda$ is  small ($\lambda\leq 10^{-4}$). More importantly, the upper bound is much lower than the outage probability achieved by all other power control schemes, which verifies that our discrete power control indeed can boost transmission capacity.

The DPC scheme in Theorem \ref{Thm:PowerAlloBoundLambda} is essentially location-dependent. Nonetheless, it can be concluded in a simple scaling form as shown in the following theorem.
\begin{theorem}\label{Thm:PowConScaLaw}
In a PCP-based ad hoc network, suppose all intended receivers in a cluster of radius $s$ are uniformly distributed. The optimal $N$-layer discrete power control that achieves the maximum contention intensity and better spatial reuse has the following scaling law
\begin{equation}\label{Eqn:DPCscaling}
P_i \in \Theta\left(\eta_i^{-\frac{\alpha}{2}}\right), \quad \forall i\in[1, 2, \ldots, N],
\end{equation}
where $\eta_i=\frac{1}{s^2}[(\sup(\mathcal{L}_i))^2-(\inf(\mathcal{L}_i))^2]$. With this power control scaling law, the cardinality of discrete power set $\mathcal{P}$ has the following scaling behavior
\begin{equation}\label{Eqn:UppScaLawN}
N\in O\left(\min_i\eta^{-\frac{\alpha}{2}}_i\right),
\end{equation}
whereas the spatial reuse factor $\delta^{\textrm{dp}}_{0_i}$ becomes
\begin{equation}\label{Eqn:ScaLawSRF}
\delta^{\textrm{dp}}_{0_i}\in\Theta\left(\frac{\delta^{\textrm{np}}_0}{N\eta_i}\right)\,\,\text{ or  }\,\,\delta^{\textrm{dp}}_{0_i}\in\Omega\left(\left(\frac{\lambda}{\eta_i}\right)^{1-\frac{2}{\alpha}}\right).
\end{equation}
\end{theorem}
\begin{IEEEproof}
See Appendix \ref{App:ProofPowConScaLaw}.
\end{IEEEproof}
\begin{remark}
Note that the power control scaling law in \eqref{Eqn:DPCscaling} and other scaling results are built based on the assumption that receivers are uniformly distributed in a cluster. If receivers are not uniformly distributed, these scaling results may not hold any more.
\end{remark}

There are several further important observations that can be concluded from Theorem \ref{Thm:PowConScaLaw} and they are specified in the following:
\begin{description}
\item[(i)] \textbf{Interference balancing}: The power control scaling law in \eqref{Eqn:DPCscaling} reflects an interesting result that a large power should be used if its selected probability is small. This intuitively makes sense since such power control balances the different interferences generated by different discrete powers and it thus reduces the total interference.
\item[(ii)] \textbf{Design of optimal discrete power control}: The power control scaling law in \eqref{Eqn:DPCscaling} can also be used to formulate an optimal discrete power design problem.  For example, consider each discrete power specified by the form $ P_i=c_i\eta_i^{-\frac{\alpha}{2}}$ where $c_i$ is a unknown constant that needs to be designed and the upper bound for $P_i$ is  $P_{\max}$. Here we choose to minimize the sum of the transmit powers $\sum_{i=1}^{N}P_i$ subject to some constraints. That is,
\begin{align}
\min\limits_{\{c_i\}} \sum_{i=1}^{N} c_i\eta_i^{-\frac{\alpha}{2}}
\end{align}
\begin{align}
\text{subject to } \sum_{j=1}^{N}c_j^{\frac{2}{\alpha}}\leq \frac{c^{\frac{2}{\alpha}}_i}{\rho_0\eta_i} , \label{Eqn:ConOutProbDPC}
\end{align}
\begin{align}
\hspace{20mm} 0< c_i\leq \eta_i^{\frac{\alpha}{2}} P_{\max}. \label{Eqn:TxPowCon}
\end{align}
where constraint \eqref{Eqn:ConOutProbDPC} is motivated by combining  $T_i\leq \kappa_{\alpha}$ and constraint \eqref{Eqn:AvgSIRIneq}, and it ensures that  the discrete power control has a lower outage probability. Constraint \eqref{Eqn:TxPowCon} is just a practical power constraint for a transmitter. This is a convex optimization problem and its solution is
\begin{equation}\label{Eqn:OptDisPowCoe}
\hspace{-.3in}c_i=\min\left\{\left(\frac{2}{\alpha}\sum_{i=1}^{N}\frac{\rho_0\eta_i-1}{2-\rho_0\eta_i}\right)^{\frac{\alpha}{\alpha-2}}\eta^{\frac{\alpha^2}{2(\alpha-2)}}_i,\eta^{\frac{\alpha}{2}}_i\right\} P_{\max},
\end{equation}
$\quad i\in\{1, 2, \ldots, N\}$. Thus the optimal discrete power control is given by
\begin{equation}\label{Eqn:OptDisPow}
P^*_i=\min\left\{\left(\frac{2\eta_i}{\alpha}\sum_{i=1}^{N}\frac{\rho_0\eta_i-1}{2-\rho_0\eta_i}\right)^{\frac{\alpha}{\alpha-2}}, 1\right\}P_{\max},
\end{equation}
$i\in\{1, 2, \ldots, N\}$.
\item[(iii)] \textbf{The optimal cardinality of power set $\mathcal{P}$}: The scaling result of the upper bound on $N$ in \eqref{Eqn:UppScaLawN} provides us a clue about how large $N$ should be. In addition, according to the proof of Theorem \ref{Thm:PowerAlloBoundLambda}, minimizing $T_i/\kappa_{\alpha}$ is roughly equivalent to minimizing the outage probability $q_i$ since both upper and lower bounds of $q_i$ are a monotonically increasing function of $T_i/\kappa_{\alpha}$. Since $T_i/\kappa_{\alpha}=\sum_{j=1}^{N}\eta_j\left(\frac{P_j}{P_i}\right)^{\frac{2}{\alpha}}\leq 1$ for all $i\in\{1, 2, \ldots, N\}$, $\sum_{j=1}^{N}\eta_j=1$ and $T_i/\kappa_{\alpha}=1$ for $N=1$, the optimal value of $N$, denoted by $N^*$, can be found by minimizing the average outage probability $\sum_{i=1}^{N}\eta_i q_i$ subject to $\sum_{i=1}^{N}\eta_i=1$. Since this objective function $\sum_{i=1}^{N}\eta_iq_i$ is too complex to be effectively handled, we can instead use $\sum_{i=1}^{N}\eta_i\frac{T_i}{\kappa_{\alpha}}[(\sup(\mathcal{L}_i))^2+(\inf(\mathcal{L}_i))^2]$ since the (tight) upper bound on $q_i$ is dominated by $T_i[(\sup(\mathcal{L}_i))^2+(\inf(\mathcal{L}_i))^2]$, i.e., finding $N^*$ by solving the following optimization problem:
\begin{eqnarray*}
&&\hspace{-.45in}\min\limits_{N} \left(\sum_{j=1}^{N}c_j^{\frac{2}{\alpha}}\right)\left(\sum_{i=1}^{N}\frac{\eta^2_i}{c_i^{\frac{2}{\alpha}}} \left[(\sup(\mathcal{L}_i))^2+(\inf(\mathcal{L}_i))^2\right] \right)\label{Eqn:OptProbN} \\
&& \text{subject to }\sum_{i=1}^{N}\eta_i=1.
\end{eqnarray*}
Since $\{\sup(\mathcal{L}_i)\}$, $\{\inf(\mathcal{L}_i)\}$ and $\{\eta_i\}$ can be determined by a predesignated cluster-partitioning rule and a given $N$ and $\{c_i\}$ can be obtained by substituting the value of $N$ and $\{\eta_i\}$ into \eqref{Eqn:OptDisPowCoe}, all variables in this optimization problem can be determined for a given $N$. Thus $N^*$ can be carried out by  searching the positive integer that minimizes the objective (cost) function in the above optimization problem. 
\end{description}

\section{Simulation Examples of $N$-Layer Discrete Power Control}
In this section, we will study two cases of $N$-layer discrete power control. First, a simple single-intended-receiver scenario is considered. Namely, each transmitter in its cluster only has one intended receiver that is distributed in $N$ different locations with certain probabilities. Next, we consider the scenario of a transmitter having multiple intended receivers. That is, each transmitter has a random number of intended receivers in a cluster that also form a homogeneous PPP. The objective of investigating these two cases is to demonstrate that our DPC scheme significantly outperforms other power control schemes already proposed in Poisson-distributed ad hoc networks.

\subsection{Single Intended Receiver with $N$ Random Locations in a Cluster}
Consider that each transmitter has only one intended receiver in a cluster and the random distance $R$ between the transmitter and the receiver is taking one of $N$ discrete values in the set $\{r_1,r_2,\ldots,r_N\}$ with the probability mass function $\mathbb{P}[R=r_i]= \eta_i$. Without loss of generality, we assume that $r_1 < r_2 < \cdots < r_N<s$.  The receivers with distance $r_i$ away from their transmitter are also called the layer-$i$ receivers. In other words, the transmitters with the receivers at layer $\mathcal{L}_i$ all have the same transmission distance  $r_i$. At each time slot, the transmitter uses power $P_i$ if its intended receiver is located in layer $i$.With the discrete power set $\mathcal{P}$, the outage probability associated with the layer-$i$ receiver is
\begin{eqnarray}
\label{Eqn:OutProbFixDis}
q_i = 1-\exp \left(-\lambda T_i \beta^{\frac{2}{\alpha}} r_i^2  \right),
\end{eqnarray}
which is easily obtained by considering a deterministic $R=r_i$ in \eqref{Eqn:OutProbLayer_i}. The optimal power control scheme that achieves the maximum contention intensity and transmission capacity is shown in the following theorem.
\begin{theorem}
\label{Thm:MainResFixDisLoc}
Suppose an intended receiver in a cluster has $N$ discrete random locations  $\{r_1, r_2,\ldots, r_N\}$ and each location $r_i$ has a corresponding probability $\eta_i$. Then the following maximum contention intensity
\begin{eqnarray}
\label{Eqn:UppBonMaxConIntFixDisLoc}
\lambda^{\textrm{dp}}_{\epsilon}  = \frac{-\log{\left(1-\epsilon\right) }}{ \kappa_{\alpha}\beta^{\frac{2}{\alpha}}\sum_{i=1}^{N}\eta_i r_i^2}.
\end{eqnarray}
is achieved with the following optimal discrete power control
\begin{eqnarray}
\label{Eqn:OptPowConFixNdis}
\frac{P_j}{P_i}=\left(\frac{r_j}{r_i}\right)^{\alpha},\,\, \text{for all }i\neq j.
\end{eqnarray}
The corresponding transmission capacity is given by
\begin{eqnarray}
\label{eq:disc+pc+TC}
C_{\epsilon}^{\textrm{dp}}= \frac{-\gamma\left(1-\epsilon\right)\log{\left(1-\epsilon\right)}}{\kappa_{\alpha}\beta^{\frac{2}{\alpha}}\sum_{i=1}^{N}\eta_i r_i^2},
\end{eqnarray}
and it is strictly greater than the transmission capacity of no power control if
\begin{eqnarray}
\label{Eqn:ConHigTCbyDPC}
\frac{r_N^2}{\sum_{i=1}^{N}{\eta_i r_i^2}} > \sum_{i=1}^{N}{\eta_i\left(1-\epsilon\right)^{\frac{r_i^2}{r_N^2}-1}}.
\end{eqnarray}
\end{theorem}
\begin{IEEEproof}
See Appendix \ref{App:ProofMainResFixDisLoc}.
\end{IEEEproof}

Note that the optimal DPC in \eqref{Eqn:OptPowConFixNdis} is equivalent to \eqref{Eqn:DPCscaling} if $\eta_i=\frac{r_i^2}{\sum_{j=1}^{N}r_j^2}$ and $\frac{P_i}{r^{\alpha}_i}=(\sum_{i=1}^{N}r_i^2)^{\frac{\alpha}{2}}$. Thus the optimal power control \eqref{Eqn:OptPowConFixNdis}  only depends on the receiver locations provided that probabilities $\{\eta_i\}$ are independent of the receiver locations. Discrete power control has the benefit of increasing the maximum contention intensity \eqref{Eqn:UppBonMaxConIntFixDisLoc} since the term $\sum_{i=1}^{N}\eta_ir_i^2$ is always smaller than $r^2_N$. The condition $\sum_{i=1}^{N}\eta_i\left(\frac{r_i}{r_N}\right)^2<1$ actually corresponds to the condition of having a lower outage probability mentioned in Remark 1. The condition of improving TC for discrete power control is given in \eqref{Eqn:ConHigTCbyDPC} and for small $\epsilon$ it can be reduced to $\sum_{i=1}^{N}\eta_i\left(\frac{r^2_i}{r^2_N}\right) \lesssim 1$, which always holds. Hence, the optimal discrete control scheme in  \eqref{Eqn:OptPowConFixNdis} is able to increase the transmission capacity for small $\epsilon$.

\begin{figure}[!t]
\centering
\includegraphics[width=3.6in,height=2.75in]{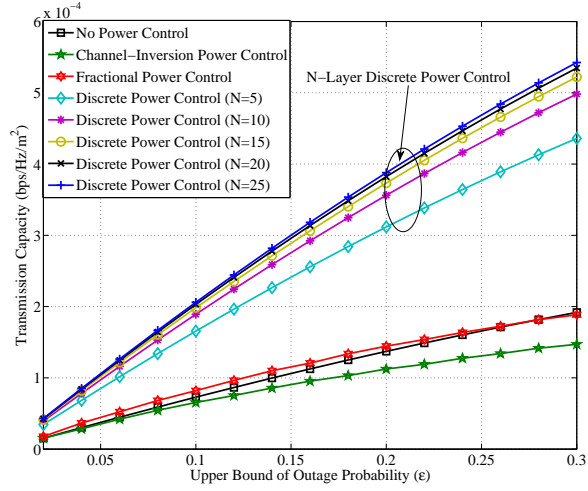}
\caption{The simulation results of transmission capacities for different power control schemes. The network parameters for simulation are $\alpha =3.5$, $\beta =1$, $s=15$m and the intended receiver is equally likely at 3m, 6m, 9m, 12m, and 15m away from its transmitter.}
\label{fig:TCPC2}
\end{figure}

\begin{figure}[!t]
\centering
\includegraphics[width=3.6in,height=2.75in]{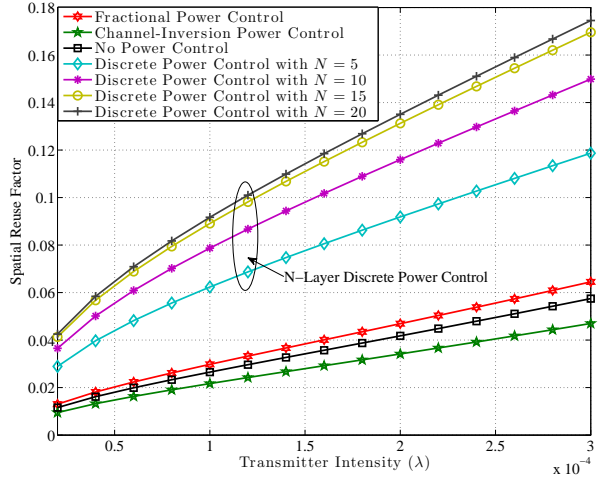}
\caption{The simulation results of spatial reuse factors for different power control schemes. The network parameters for simulation is $\alpha =3.5$, $s=15$m and $\eta_i=\frac{1}{5}$. The intended receivers of a transmitter are equally likely located at 3m, 6m, 9m, 12m, and 15m away from their transmitter.}
\label{fig:SpaReuFac2}
\end{figure}

We now present some numerical simulations regarding the results in Theorem \ref{Thm:MainResFixDisLoc} by assuming $r_i=\frac{i}{N}s$ and $\eta_i=\frac{1}{N}$ for all $i$. Under this assumption, we have $\frac{P_i}{P_j}=\frac{i^{\alpha}}{j^{\alpha}}$ and $\sum_{i=1}^{N}\eta_i(\frac{r_i}{r_N})^2=\frac{1}{6}(1+\frac{1}{N})(2+\frac{1}{N})$. Thus the condition in \eqref{Eqn:ConHigTCbyDPC} for small $\epsilon$ can be simplified to
$\sum_{i=1}^{N}\eta_i\left(\frac{r^2_i}{r^2_N}\right)=\frac{(1+1/N)(2+1/N)}{6}$, which is always smaller than one and approaches its minimum at $\frac{1}{3}$ as $N$ gets large. This means that the discrete power control under this setting can achieve nearly 3 times transmission capacity than no power control if $N$ is large and $\epsilon$ is small. However, a super large $N$ is not proper in practice and it has a diminishing return problem.  Fig. \ref{fig:TCPC2} illustrates  the effects of the proposed optimal discrete power control on enhancing the transmission capacity when the radius of a cluster $s$ is 15m and it is segmented to $5$ equal lengths of 3m, i.e., $r_i=3i$ and $\eta_i=\frac{1}{5}$. The transmission capacities achieved by no power control, channel-inversion, and fractional power control schemes  are also illustrated for comparison. The transmit powers for each transmitter $X_i$ using fractional power control and each transmitter $X_j$ using channel-inversion power control are $1/\sqrt{H_i}$ and $1/H_j$, respectively. As we see,  $N$-layer discrete power control significantly outperforms all other power control schemes in terms of transmission capacity, and increasing $N$ can increase TC. However, using a large $N$ does not produce too much benefit on TC and it looks like $N=15$ is good enough in this case. Similar observations can also be acquired from the simulation results of the spatial reuse factors in Fig. \ref{fig:SpaReuFac2}.

\subsection{Multiple Intended Receivers Uniformly Distributed in a Cluster}

Now we investigate and simulate the case that each transmitter has multiple intended receivers uniformly distributed in its cluster with $N$ layers, and at each time slot the transmitter independently selects one of the intended receivers to transmit with probability $\eta_i$  if the selected receiver is at the $i$th layer. Reference \cite{CHLJGA11} showed that the selected receivers also form a homogeneous PPP of intensity $\lambda$. Each cluster is layered by segmenting the cluster radius $s$ into $N$ equal lengths of $\frac{s}{N}$, such that the $i$th layer is the annulus with inner radius of $\frac{(i-1)s}{N}$ and outer radius of $\frac{is}{N}$, and thus  the probability of the selected receiver being in the $i$th layer is $\eta_i=\frac{2i-1}{N^2}$. Note that $\eta_i$ increases along its index $i$ such that the intended receivers in a farther layer can be selected for service more often. According to the discrete optimal power in Theorem \ref{Thm:PowerAlloBoundLambda}, we know that the following power ratio
\begin{equation}
\frac{P_j}{P_i}=\left[\frac{(2j-1)(j^2+(j-1)^2)}{(2i-1)(i^2+(i-1)^2)}\right]^{\frac{\alpha}{2}}\left(\frac{\eta_i}{\eta_j}\right)^{\frac{\alpha}{2}}
\end{equation}
can achieve the following lower bound on TC
\begin{equation}
 \underline{C}^{\textrm{dp}}_{\epsilon}= \frac{2\gamma \epsilon (1-\epsilon)}{\kappa_{\alpha}\beta^{\frac{2}{\alpha}}s^2}.
\end{equation}
Also, the following power ratio $$\frac{P_j}{P_i}=\left(\frac{j-1}{i-1}\right)^{\alpha}=\left[\frac{(2j-1)(j-1)^2}{(2i-1)(i-1)^2}\right]^{\frac{\alpha}{2}}\left(\frac{\eta_i}{\eta_j}\right)^{\frac{\alpha}{2}}$$
for $i, j\neq 1$ and $\inf(\mathcal{L}_1)>0$ can achieve the following upper bound on TC
\begin{eqnarray}
\overline{C}^{\textrm{dp}}_{\epsilon} =  \frac{2\gamma  \epsilon}{\kappa_{\alpha}\beta^{\frac{2}{\alpha}}s^2 \left(1-\frac{4}{3N}+\frac{1}{3N^3}\right)},\,\,N>1.
\end{eqnarray}
Thus, we can choose $P_i=c_i\eta_i^{-\frac{\alpha}{2}}$ where $(2i-1)(i-1)^2<c^{\frac{2}{\alpha}}_i< (2i-1)(i^2+(i-1)^2)$. Note that the lower bound $\underline{C}^{\textrm{dp}}_{\epsilon}$ is exactly twice of the TC achieved by no power control for small $\epsilon$, which certainly indicates that using discrete power control can achieve a larger TC than no power control\footnote{The transmit power for no power control (and others) is always set according to the worse-case transmission distance $s$ since there is always a possibility that an intended receiver is located at $s$.}. In addition, comparing $\overline{C}_{\epsilon}^{\textrm{dp}}$ with $\underline{C}^{\textrm{dp}}_{\epsilon}$ reveals that  using a very large $N$ should be avoided since $\overline{C}_{\epsilon}^{\textrm{dp}}\approx \underline{C}^{\textrm{dp}}_{\epsilon}$ as $N\gg 1$ and $\epsilon\ll 1$, and $\overline{C}_{\epsilon}^{\textrm{dp}}$ is maximized and 5 times more than the TC of no power control when $N=2$.

\begin{figure}[!t]
\centering
\includegraphics[width=3.6in,height=2.75in]{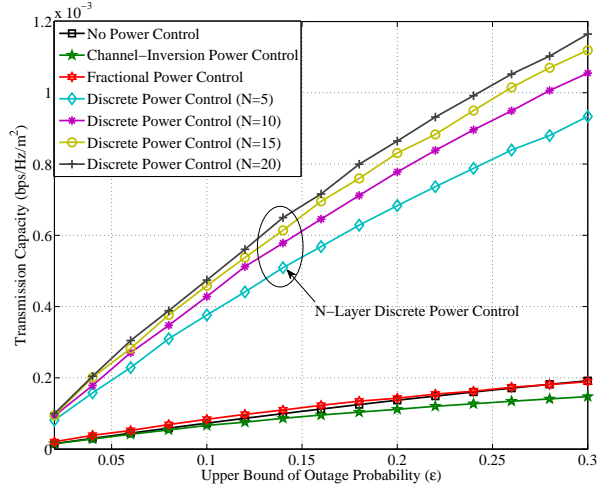}
\caption{The simulation results of transmission capacities for different power control schemes. The network parameters for simulation are $s=15$m, $\alpha =3.5$ and $\beta =1$. The intended receivers of a transmitter are uniformly distributed in a cluster.}
\label{fig:TCPC3}
\end{figure}

\begin{figure}[!t]
\centering
\includegraphics[width=3.6in,height=2.75in]{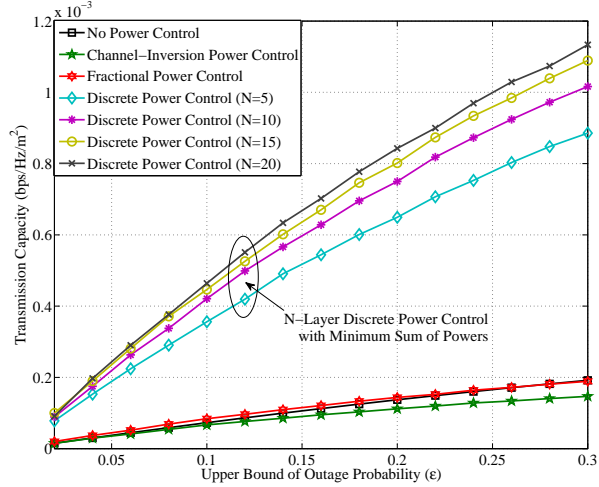}
\caption{The simulation results of transmission capacities for optimal discrete power control  and other power control schemes. The network parameters for simulation are $s=15$m, $\alpha =3.5$ and $\beta =1$. The intended receivers of a transmitter are uniformly distributed in a cluster.}
\label{fig:TCPC4}
\end{figure}

The simulation results of transmission capacities for discrete power control $P_i=c_i\eta^{-\frac{\alpha}{2}}_i$ with parameters $\eta_i=\frac{2i-1}{N^2}$ and $c_i=\left(\frac{3}{2}(2i-1)(i-1)^2\right)^{\frac{\alpha}{2}}$ and different values of $N$ are shown in Fig.~\ref{fig:TCPC3}. As expected, all $N$-layer discrete power controls can achieve at least twice TC of other control schemes, and a higher value of $N$ can lead to a higher TC. The maximum of TC for $N$-layer power control will be attained when $N$ goes to infinity; however, Fig. \ref{fig:TCPC3} shows that $N=20$ is good enough for approaching the maximum TC. Fig. \ref{fig:TCPC4} shows the simulation results of the optimal discrete power scheme in \eqref{Eqn:OptDisPow} with $P_{\max}=1$ with the same network parameters as used in Fig. \ref{fig:TCPC3}. The transmission capacities for different values of $N$ in Figs. \ref{fig:TCPC4} are very much similar to those in Fig. \ref{fig:TCPC3}, but the sum powers used in Fig. \ref{fig:TCPC4} is just about $75\%\sim 80\%$ of the sum of the discrete powers used in Fig. \ref{fig:TCPC3}. Thus using \eqref{Eqn:OptDisPow} is able to reduce the power cost while keeping the same level of the TC performance.


\section{Conclusion}
The $N$-layer DPC scheme proposed in this paper is mainly motivated by the fact that practical power control in a digital device is of a discrete nature. We first show that in a Poisson-distributed network, a discrete power control is able to work  strictly better than no power control in the sense of a mean SIR and outage probability, provided that some constrains on the discrete powers and their selected probabilities are satisfied. In particular, we design an $N$-layer DPC scheme in which $N$ discrete powers can be used by the transmitters, where which discrete power is used depends on which layer the intended receiver is located in a cluster with $N$ layers. In order to evaluate the performance of the proposed discrete power control, the transmission capacity and outage-free spatial reuse factor are redefined. The average outage probability of each layer is derived, which is the foundation of developing the optimal discrete power control scaling law $P_i=\Theta\left(\eta^{-\frac{\alpha}{2}}\right)$. The optimization methods of choosing the discrete power and the cardinality of the power set are also discussed. Finally, two simulation examples are presented to show that the proposed $N$-layer discrete power control is able to achieve larger transmission capacity and spatial reuse factor than no-power control and other existing power control schemes.

\appendix[Proofs of Theorems]
\subsection{Proof of Theorem \ref{Thm:AvgSIRInq}}\label{App:ProofAvgSIRInq}
First consider the case that each transmitter uses a single transmit power $P_0$. The average SIR in \eqref{Eqn:DefnSIR} for this case can be modified as
$$\mathbb{E}[\mathrm{SIR}_0] = \mathbb{E}\left[ R^{-\alpha}\right]\mathbb{E}\left[I^{-1}_0(1)\right].$$
Now consider that each transmitter has $N$ discrete power levels. Then the average $\mathrm{SIR}_0$ in \eqref{Eqn:DefnSIR} with transmit power $P_i$ can be found as
$$\mathbb{E}[\mathrm{SIR}_{0_i}] = P_i\mathbb{E}\left[ R^{-\alpha}\right]\mathbb{E}\left[\frac{1}{\sum_{j=1}^N I_j}\right],\,\,i\in\{1, 2, \ldots, N\},$$
where $I_j $ is given by
\begin{eqnarray*}
I_j=P_j \sum_{X_k\in\Phi_j\setminus X_0}H_{k0}\|X_k\|^{-\alpha},\,\,j\in\{1, 2, \ldots, N\},
\end{eqnarray*}
in which $\Phi_j$ is a homogeneous PPP with intensity $\eta_j\lambda$ that consists of transmitters independently selecting power $P_j$. According to the conservation property shown in Lemma \ref{Lam:ConserversionPropertyPCP}, $\Phi_j$ can be transformed into another homogeneous PPP $\Phi'_j$ of intensity $\lambda$. Let $I_j'$ denote the interference at the origin generated by $\Phi'_j$ and it is given by
$$I'_j=P_j \sum_{X'_k\in\Phi_j'\setminus X_0}H_{k0}\|X'_k\|^{-\alpha},$$
where $X'_k=\eta^{\frac{1}{2}}_jX_k$ for all $X_k\in\Phi_j$. Since $\eta_j^{\frac{\alpha}{2}}I'_j$ can be viewed as the interference generated at the origin by a homogeneous Poisson-distributed transmitters with intensity $\eta_j\lambda$ and power $P_j$, $\eta_j^{\frac{\alpha}{2}}I'_j$ and $I_j$ are both generated by a homogeneous PPP of intensity $\eta_j\lambda$. Accordingly, we know 
\begin{align*}
\eta_j^{\frac{\alpha}{2}}I'_j &= P_j\sum_{X'_k\in\Phi'_j\setminus X_0} H_{k0}\|\eta_j^{-\frac{1}{2}}X'_k\|^{-\alpha}\\
&\stackrel{d}{=} P_j\sum_{X_k\in\Phi_j\setminus X_0} H_{k0}\|X_k\|^{-\alpha}=I_j\\
&\stackrel{d}{=}\eta_j^{\frac{\alpha}{2}}P_jI_0(1),
\end{align*}
where $\stackrel{d}{=}$ stands for equivalence in distribution. So it turns out that $\mathbb{E}[I_j]=\eta_j^{\frac{\alpha}{2}}P_j\mathbb{E}[I_0(1)]$.

By Jensen's inequality, $\mathbb{E}[\mathrm{SIR}_{0i}]$ can be lower-bounded as
$$\mathbb{E}[\mathrm{SIR}_{0_i}]\geq \frac{P_i\mathbb{E}[R^{-\alpha}]}{\sum_{j=1}^{N}\mathbb{E}[I_j]}=\frac{P_i\mathbb{E}[R^{-\alpha}]}{\mathbb{E}[I_0(1)]\sum_{j=1}^{N}\eta_j^{\frac{\alpha}{2}}P_j},$$
which further gives
$$\mathbb{E}[\mathrm{SIR}_{0_i}]\geq \mathbb{E}[R^{-\alpha}]\mathbb{E}[I_0^{-1}(1)]$$
based on \eqref{Eqn:AvgSIRIneq}. Thus, $\mathbb{E}[\mathrm{SIR}_{0_i}]\geq \mathbb{E}[\mathrm{SIR}_0]$. Also, we know
\begin{align*}
\hspace{-.3in}\mathbb{E}[\mathrm{SIR}_{0_i}]-\mathbb{E}[\mathrm{SIR}_0]&= \int_{0}^{\infty}(\mathbb{P}[\mathrm{SIR}_0\leq \beta]\\
&\hspace{.35in}-\mathbb{P}[\mathrm{SIR}_{0_i}\leq \beta])\,\, \dif \beta  \geq 0,
\end{align*}
and $\mathbb{P}[\mathrm{SIR}\geq \beta]$ is a monotonic function of $\beta$. Thus, we must have $\mathbb{P}[\mathrm{SIR}_{0_i}< \beta]\leq \mathbb{P}[\mathrm{SIR}_0< \beta]$, which means that any transmitter using more-than-one discrete powers does not have a higher outage probability.

\subsection{Proof of Theorem \ref{Thm:OutProbLayi}}\label{App:ProofOutProbLayi}
First we point out that transmitters using power $P_i$ form a homogeneous PPP with intensity $ \eta_i\lambda$. This is due to the fact that a receiver is located at layer-$i$ with probability $\mathbb{P}[R\in\mathcal{L}_i] =\eta_i$ and the distances of receivers are i.i.d. across different clusters. Also, each transmitter independently selects its own transmit power and thus the resulting process of transmitters with transmit power $P_i$ forms a thinned homogeneous PPP ($\Phi_i$) with intensity $\lambda_i = \eta_i\lambda$. Thus all thinned transmitter point processes are mutually independent, i.e., $\Phi_i$ is independent of $\Phi_j$ for $i \neq j$, and $\Phi = \bigcup_{j=1}^{N}{\Phi_j}$. For a layer-$i$ receiver located at distance $R_i\in \mathcal{L}_i$ away from its transmitter, $\mathrm{SIR}_0(P_i)$ in \eqref{Eqn:DefnSIR} becomes
\begin{align}\label{eq:continous+SIR}
\mathrm{SIR}_0(P_i) &= \frac{P_i H_0 R_i^{-\alpha}}{  \sum_{j=1}^{N}{\sum_{X_{j_k} \in \Phi_j \setminus X_0}{P_j \tilde{H}_{j0}   ||X_{j_k} - Y_0 ||^{-\alpha}}  } }\nonumber\\
&\stackrel{d}{=}\frac{P_i H_0 R_i^{-\alpha}}{\sum_{j=1}^{N}I_0(\eta_j^{\frac{\alpha}{2}}P_j)}.
\end{align}
The outage probability at a given $R_i=r$ can be calculated as
\begin{align}
q_{i}(r) &= 1-\mathbb{P}\left[ \mathrm{SIR}_0(P_i)> \beta\bigg|R_i=r\right]\nonumber\\
&= 1-\mathbb{P}\left[H_0 > \frac{\beta r^\alpha}{P_i}\sum_{j=1}^{N}I_0(\eta^{\frac{\alpha}{2}}P_j) \right]\nonumber\\
&= 1- \mathbb{E}\left[e^{- \frac{\beta r^\alpha}{P_i}\sum_{j=1}^{N}I_0(\eta_j^{\frac{\alpha}{2}}P_j)}\right]\nonumber\\
&= 1- \prod_{j=1}^{N}{\mathbb{E}\left[e^{- \frac{\beta r^\alpha}{P_i} I_0(\eta^{\frac{\alpha}{2}}P_j) }\right] } \nonumber \\
 &\stackrel{(b)}{=}  1-\exp \left\{ -\lambda\kappa_{\alpha} \beta^{\frac{2}{\alpha}} r^{2} \left[ \sum_{j=1}^{N}\eta_j \left(\frac{P_j}{P_i}\right)^{\frac{2}{\alpha}} \right] \right\}    \nonumber \\
 &= 1 - \exp \left( -\lambda \beta^{\frac{2}{\alpha}} r^{2}T_i\right),\label{eq:cont+outage+r}
\end{align}
where  $(b)$ follows from the outage expression for Rayleigh fading without noise (see (16.8) in \cite{Baccllibook}).

Let $F_{R_i}(r)$ and $f_{R_i}(r)$ denote the cdf and pdf of the random distance $R_i$ in layer $i$, respectively. Conditioning on that the receiver distance to its transmitter falls into $\mathcal{L}_i$, the probability that an intended receiver is located at $R_i\leq r \in \mathcal{L}_i$ is given by
\begin{align}
F_{R_i}(r) &= \mathbb{P}[R_i\leq  r|R_i\in\mathcal{L}_i] =\frac{\mathbb{P}[R_i\in\mathcal{B}(0,r)\cap\mathcal{L}_i]}{ \mathbb{P}[R_i\in\mathcal{L}_i]}\nonumber\\
 &=\frac{ F_R(r) - F_R(\sup\{\mathcal{L}_i\})}{\eta_i},
\end{align}
where $\mathcal{B}(0,r)$ represents a circular disk of radius $r$ located at the origin. As a result, the pdf $f_{R_i}(r)$ can be shown as $f_{R_i}(r) = \frac{1}{\eta_i}f_R(r)$. Therefore, the outage probability of the layer-$i$ receiver is
\begin{align*}
q_i =\mathbb{E}_{R_i}[q_i(R_i)] &= 1-\frac{1}{\eta_i}\int_{\mathcal{L}_i} e^{-\lambda\beta^{\frac{2}{\alpha}}r^2T_i}f_R(r)\dif r \nonumber\\
&= 1-\mathbb{E}\left[ e^{-\lambda\beta^{\frac{2}{\alpha}}R^2T_i}\bigg|R\in\mathcal{L}_i\right].
\end{align*}
Thus \eqref{Eqn:OutProbLayer_i} is attained.


\subsection{Proof of Theorem \ref{Thm:PowConScaLaw}}\label{App:ProofPowConScaLaw}
According to the power ratio result in \eqref{eq:cont+lower+power}, we have
\begin{align}
\frac{P_j}{P_i} &= \left[\frac{(\sup(\mathcal{L}_j))^2+(\inf(\mathcal{L}_j))^2}{(\sup(\mathcal{L}_i))^2+(\inf(\mathcal{L}_i))^2}\right]^{\frac{\alpha}{2}}\nonumber\\
&= \left(\frac{\eta_j}{\eta_i}\right)^{-\frac{\alpha}{2}}\left[\frac{(\sup(\mathcal{L}_j))^4-(\inf(\mathcal{L}_j))^4}{(\sup(\mathcal{L}_i))^4-(\inf(\mathcal{L}_i))^4}\right]^{\frac{\alpha}{2}}.
\end{align}
If $\sup(\mathcal{L}_j)\leq \sup(\mathcal{L}_i)$,  then there must exist a constant $M_1>0$ such that
$$\frac{P_j}{P_i}\geq  \left(\frac{\eta_j}{\eta_i}\right)^{-\frac{\alpha}{2}}M_1$$
since $\sup(\mathcal{L}_i), \sup(\mathcal{L}_j)\in O(s)$. On the contrary, if  $\sup(\mathcal{L}_j)\geq \sup(\mathcal{L}_i)$, then there must also exist a constant $M_2>0$ such that $$\frac{P_j}{P_i}\leq \left(\frac{\eta_j}{\eta_i}\right)^{-\frac{\alpha}{2}}M_2$$
since $\inf(\mathcal{L}_i), \inf(\mathcal{L}_j)\in\Omega(s)$.

From \eqref{eq:cont+upper+power}, we also can have
\begin{align}
\frac{P_j}{P_i}&= \left[\frac{(\inf(\mathcal{L}_j))^2}{(\inf(\mathcal{L}_i))^2}\right]^{\frac{\alpha}{2}} \geq  \left[\frac{(\sup(\mathcal{L}_j))^2+(\inf(\mathcal{L}_j))^2}{(\sup(\mathcal{L}_i))^2+(\inf(\mathcal{L}_i))^2}\right]^{\frac{\alpha}{2}}\nonumber\\
&\hspace{.2in}\cdot\left[\frac{(\sup(\mathcal{L}_i))^2+(\inf(\mathcal{L}_i))^2}{(\sup(\mathcal{L}_j))^2+(\inf(\mathcal{L}_i))^2}\right]^{\frac{\alpha}{2}} \geq  M_1\left(\frac{\eta_j}{\eta_i}\right)^{-\frac{\alpha}{2}}
\end{align}
if $\inf(\mathcal{L}_j)\geq\inf(\mathcal{L}_i)$. Similarly, if $\inf(\mathcal{L}_j)\leq\inf(\mathcal{L}_i)$ then we can show
\begin{equation}
\frac{P_j}{P_i}\leq M_2\left(\frac{\eta_j}{\eta_i}\right)^{-\frac{\alpha}{2}}.
\end{equation}
Since the power ratios $\frac{P_j}{P_i}$ for achieving the upper and lower bounds on the maximum contention intensity obey the scaling law of $\left(\frac{\eta_j}{\eta_i}\right)^{-\frac{\alpha}{2}}$, it follows that $P_i\in\Theta\left(\eta_i^{-\frac{\alpha}{2}}\right)$ achieves the maximum contention intensity. In addition, it is easy to verify that $\frac{P_j}{P_i}\in\Theta\left(\left(\frac{\eta_j}{\eta_i}\right)^{-\frac{\alpha}{2}}\right)$ satisfies \eqref{Eqn:AvgSIRIneq} and thus $P_i\in\Theta\left(\eta_i^{-\frac{\alpha}{2}}\right)$ achieves better spatial reuse. Finally, substituting this power control scaling law into \eqref{Eqn:AvgSIRIneq} and $\delta_{0_i}$ in Lemma \ref{Lem:SpatialReuse} gives us the scaling laws of upper and lower bounds on $N$ and $\delta_{0_i}$.

\subsection{Proof of Theorem \ref{Thm:MainResFixDisLoc}}\label{App:ProofMainResFixDisLoc}
Recall that the outage probability of each layer is upper-bounded by $\epsilon$, i.e., $\max_i q_i \leq \epsilon$ as given in \eqref{eq:def+OP}. According to the proof of Theorem \ref{Thm:PowerAlloBoundLambda}, the maximum contention intensity $ \overline{\lambda}_\epsilon $ is achieved when $\{q_i\}$ in \eqref{Eqn:OutProbFixDis} for all $i$ are the same and this yields
\begin{equation}
\frac{P_1}{r_1^{\alpha}} = \frac{P_2}{r_2^{\alpha}} = \cdots =\frac{P_N}{r_N^{\alpha}}.
\end{equation}
By substituting it into the outage probabilities, the maximum contention intensity can be acquired as given in \eqref{Eqn:UppBonMaxConIntFixDisLoc}. Under the optimal power control scheme in \eqref{Eqn:OptPowConFixNdis}, all receivers at different locations undergo the same outage probabilities with $q_{i}(\overline{\lambda}_{\epsilon}^{\textrm{dp}})= \epsilon$; therefore, the transmission capacity can be derived based on the definition as
\begin{eqnarray}
C_{\epsilon}^{\textrm{dp}} = \gamma(1-\epsilon) \overline{\lambda}_{\epsilon}^{\textrm{dp}} =  \frac{-\gamma\left(1-\epsilon\right)\log{\left(1-\epsilon\right)}}{\kappa_{\alpha}\beta^{\frac{2}{\alpha}}\sum_{i=1}^{N}\eta_ir_i^2}.
\end{eqnarray}

If there is no power control, the SIR at the layer-$i$ receiver is
\begin{eqnarray}
\mbox{SIR}_{0_i}^{\textrm{np}} = \frac{ H_0 r_{i}^{-\alpha}}{ \sum_{ X_j \in \Phi \setminus X_0 }{H_{ji} ||X_j - r_i ||^{-\alpha} }}.
\end{eqnarray}
According to (16.8) in \cite{Baccllibook}, the outage probability associated with layer-$i$ receiver can be found as
\begin{eqnarray}
q_i^{\textrm{np}} = 1-\exp\left\{ -\lambda \kappa_{\alpha}\beta^{\frac{2}{\alpha}} r_i^2 \right\}, \;\;\;\; i \in [1, 2, \ldots, N].
\end{eqnarray}
Since $r_1 < r_2 < \cdots  <r_{i} <\cdots < r_N<s$, the largest outage probability is $q_{N}^{\textrm{np}}$, which is the outage probability of the layer-$N$ receiver. Therefore, the maximum contention intensity that satisfies the outage probability constraint $q_{N}^{\textrm{np}}= \epsilon$ is given by
\begin{eqnarray}
\label{eq:disc+contention+npc}
 \overline{\lambda}_{\epsilon}^{\textrm{np}} = \frac{-\log{\left(1-\epsilon\right)}}{\kappa_{\alpha}\beta^{\frac{2}{\alpha}} r_N^2 }.
\end{eqnarray}
By comparing \eqref{Eqn:UppBonMaxConIntFixDisLoc} with \eqref{eq:disc+contention+npc} and using the fact $\sum_{i=1}^{N}{\eta_i r_i^2} < r_N^{2}$, we have   $ \overline{\lambda}_{\epsilon}^{\textrm{dp}} > \overline{\lambda}_{\epsilon}^{\textrm{np}}$; namely, discrete power control increases the maximum contention intensity. Thus the maximum allowable transmitter intensity is now equal to $\overline{\lambda}_{\epsilon}^{\textrm{np}}$, and the outage probability $q_i^{\textrm{np}}$ with $\overline{\lambda}_{\epsilon}^{\textrm{np}}$ becomes $q_{i}^{\textrm{np}}= 1-\exp\left\{ -\overline{\lambda}^{\textrm{np}}_{\epsilon} \kappa_{\alpha}\beta^{\frac{2}{\alpha}} r_i^2 \right\} = 1- \left(1-\epsilon \right)^{r_i^2/r_N^2}$. The resulting transmission capacity is given by
\begin{align}
\label{Eqn:NoPowConTCwFixDisLoc}
C_{\epsilon}^{\textrm{np}}&=  \gamma\overline{\lambda}_{\epsilon}^{\textrm{np}}  \sum_{i=1}^{N}\eta_i\left( 1- q_i^{\textrm{np}}\left( \overline{\lambda}_{\epsilon}^{\textrm{np}}  \right)\right)\nonumber \\
&= \frac{-\gamma\log{\left(1-\epsilon\right)}}{\beta^{\frac{2}{\alpha}}\kappa_{\alpha}r_N^2 }\sum_{i=1}^{N}\eta_i(1-\epsilon)^{\frac{r_i^2}{r_N^2}}.
\end{align}
Therefore, the transmission capacity with $N$-layer discrete power control in \eqref{eq:disc+pc+TC} is larger than that in \eqref{Eqn:NoPowConTCwFixDisLoc} if the condition in \eqref{Eqn:ConHigTCbyDPC} holds.


\bibliographystyle{ieeetran}
\bibliography{IEEEabrv,Ref_DDBPC}

\end{document}